\documentclass[pra,twocolumn,showpacs,superscriptaddress,floatfix]{revtex4-2}
\usepackage{graphicx}
\usepackage{epstopdf} 
\usepackage{amsmath}
\usepackage{epsfig}
\usepackage{latexsym}
\usepackage{amsfonts}
\usepackage{graphics}
\usepackage{bm}
\usepackage{braket}
\usepackage{dsfont}
\usepackage{soul}
\usepackage{ulem}
\usepackage{bbm}

\usepackage{mathtools}

\usepackage{helvet}

\graphicspath{ {./Graphics/} }
\begin{document}

\date{\today}
\author{W. Kirkby}
\affiliation{Department of Physics and Astronomy, McMaster University, 1280 Main St.\ W., Hamilton, ON, L8S 4M1, Canada}
\author{D. H. J. O'Dell}
\affiliation{Department of Physics and Astronomy, McMaster University, 1280 Main St.\ W., Hamilton, ON, L8S 4M1, Canada}
\author{J. Mumford}
\affiliation{Department of Physics and Astronomy, McMaster University, 1280 Main St.\ W., Hamilton, ON, L8S 4M1, Canada}
\affiliation{School of Arts and Sciences, Red Deer College, 100 College Boulevard, Red Deer, AB, T4N 5H5, Canada}

\title{False signals of chaos from quantum probes}

\begin{abstract}
We demonstrate that two-time correlation functions, which are generalizations of out-of-time-ordered correlators (OTOCs), can show `false-flags' of chaos by exhibiting behaviour predicted by random matrix theory even in a system with classically regular dynamics. In particular, we analyze a system of bosons trapped in a double-well potential and probed by a quantum dot which is coupled to the bosons dispersively.  This is an integrable system (considered both as separate parts and in total). Despite the continuous time evolution generated by the actual Hamiltonian, we find that the $n$-fold two-time correlation function for the probe  describes an effective stroboscopic or Floquet dynamics whereby the bosons appear to be alternately driven by two different non-commuting Hamiltonians in a manner reminiscent of the Trotterized time evolution that occurs in digital quantum simulation. The classical limit of this effective dynamics can have a nonzero Lyapunov exponent, while the effective level statistics and return probability show traditional signatures of chaotic behaviour. In line with several other recent studies, this work highlights the fact that the behavior of OTOCs and their generalizations must be interpreted with some care.
\end{abstract}

\maketitle

\section{\label{Sec:Intro}Introduction}

Two-time correlation (TTC) functions are indispensable tools in the investigation of the dynamics of quantum many-body systems. For example, a TTC of the form $\langle [\hat{A}(t) , \hat{B}(t')] \rangle$ enters Kubo's formula for the linear response of the observable ${A}$ at time $t$ due to the time-dependent perturbation at earlier times $t'$ by the drive $B$  \cite{chaiken_lubensky}.
The wide utility of linear response theory means that TTCs are therefore a vital ingredient in calculations in quantum many-particle kinetics ranging from absorption spectra, to reaction rates and diffusion constants. They are also useful in assessing `quantumness' through their connection to Leggett-Garg inequalities \cite{leggett08, emary14}.

In this paper we consider a general $n$-fold TTC function \cite{bhattacharyya19,haehl19,halpern18,shenker14,roberts17} which for Hermitian operators we define as  
\begin{equation}
F_n(t) \equiv \Big \langle \left[ \hat{A}(t) \hat{B}(0) \right]^{n} \Big \rangle 
\label{eq:TTC}
\end{equation}
 where $\hat{A}(t) = e^{i \hat{H} t} \hat{A}(0) e^{-i\hat{H}t}$, and $\langle \dots \rangle$ is the expectation value taken with respect to a pure or mixed state.  
 The first order TTC function $F_1 = \langle \hat{A}(t) \hat{B}(0) \rangle$ describes a perturbation by operator $\hat{B}$ at time $t = 0$ followed by a `probe' by operator $\hat{A}$ at time $t$ like in the Kubo formula.  This function 
is related to quantities such as the quantum fidelity which has been successfully employed as a means of detecting and characterizing quantum phase transitions (QPTs) \cite{zanardi06,chen07,buonsante07,quan06,ning08}.  However, in general, the first order TTC fails to capture the spread of information across a system from an initial perturbation.   Thus, in recent years the second order TTC function $F_2 = \langle \hat{A}(t)\hat{B}(0)\hat{A}(t)\hat{B}(0)\rangle$ has gained popularity and is often referred to as the out-of-time-ordered correlation (OTOC) function. 

 In addition to being more sensitive to QPTs than first order TTCs \cite{heyl18,sun19,mumford20}, OTOCs have been used to identify the `scrambling' of information across a system's degrees of freedom \cite{yao16,swingle16,bohrdt17}.   For this purpose it is useful to  express the OTOC function as an overlap between two states, $F_2(t) = \langle \psi_1 (t)\vert \psi_2(t) \rangle$ where $\vert \psi_1(t)\rangle = \hat{B}(0) \hat{A}(t) \vert \psi \rangle$ and  $\vert \psi_2(t)\rangle = \hat{A}(t) \hat{B}(0) \vert \psi \rangle$ and $\vert \psi \rangle $ is some general state.  When the operators are chosen such that they initially commute, $[ \hat{A}(0), \hat{B}(0) ] = 0$, then $F_2$ is unity at $t = 0$, and at later times it decays as correlations build up and these operators no longer commute. In complex systems it turns out that there is typically an exponential decay  $F_2(t) \approx 1 - c e^{\lambda t}$ where $c$ is some constant and $\lambda$ is the decay rate. 

The exponential sensitivity of OTOCs to information scrambling has led to the exciting idea that OTOCs might be capable of quantifying many-body quantum chaos (or stated more carefully, dynamics which would be chaotic in the classical limit) \cite{roberts15,maldacena2016a,maldacena2016b,zhu16,kukuljan17,rozenbaum17,cotler18,kurchan18,chen18,mata18,jalabert18,hamazaki18,herrera18,rozenbaum19,carlos19}. The defining feature of classical chaos is an exponential sensitivity to initial conditions, i.e.\ the exponential increase in separation over time of initially close points in phase space, and is quantified by Lyapunov exponents. This behaviour is considered to be a prerequisite for ergodicity  and thermalization which destroys any memory of the initial state and it therefore seems natural enough from an information-theoretic perspective that chaos should be related to information scrambling.  In fact, it has been demonstrated in a number of specific examples that the OTOC decay rate $\lambda$ is directly related to the Lyapunov exponent $\lambda_L$ in the classical limit of a chaotic quantum system; these cases include the kicked rotor \cite{hashimoto2017}, stadium billiard \cite{rozenbaum19},  Dicke model  \cite{carlos19} and kicked Dicke model \cite{sinha2021}. 

However, recently it has been shown that having $\lambda > 0$ for an OTOC does not necessarily indicate that the system is chaotic, but instead can be caused by information scrambling from dynamics near an unstable fixed point \cite{pappalardi18,hummel19}. The simplest example of such cases is the exponential separation of trajectories at short times in the inverted harmonic oscillator, which in single-particle quantum mechanics gives rise to $\lambda>0$ at finite and infinite temperatures from the OTOC \cite{hashimoto20}. Furthermore, OTOCs in integrable many-particle systems such as the Lipkin-Meshkov-Glick model and the Dicke model  (in the latter's integrable phase) \cite{cameo20,xu20} also exhibit positive $\lambda$ near unstable points resulting from second-order QPTs. 


In this paper we provide another example of how TTCs can exhibit false-flags of chaos by showing that for a regular (i.e.\ nonchaotic) system the operator in Eq. \eqref{eq:TTC} can display Wigner Dyson-type spectral statistics described by random matrix theory (RMT).   RMT was first used in the 1950s to understand the statistical properties of the spectra of complex nuclei \cite{wigner51} and reached maturity in the 1980s with the realization (as encapsulated by conjectures such as that due to Bohigas, Giannoni, and Schmit  \cite{bohigas1983,bohigas84}) that fluctuations in the distances between energy levels have universal properties in the semiclassical regime (far above the ground state) that distinguish chaotic from nonchaotic systems. In fact, it seems that apart from a few atypical exceptions the spectral statistics of physical systems fall into  one of four classes determined by ensembles of random Hermitian matrices. For classically integrable systems the statistics of the corresponding quantum energy levels are Poissonian, whereas for classically chaotic systems the corresponding energy level fluctuations follow those of either  the Gaussian orthogonal ensemble (GOE), the Gaussian unitary ensemble (GUE) or the Gaussian symplectic ensemble (GSE) independent of the details of the system and depending only on the symmetry properties of the Hamiltonian under canonical transformations. For a Floquet-type unitary operator [such as that effectively given by Eq.\ \eqref{eq:TTC}],  these ensembles are changed from Gaussian to circular ensembles (COE etc.) because the eigenvalues should have a magnitude of one.


The system we use to illustrate these features is a simple model consisting of $N$ identical bosons occupying two modes and coupled to a single qubit probe (atomic quantum dot) and has previously been discussed in a considerable number of theoretical proposals, e.g.\ \cite{bausmerth07,Rinck2011,mulansky11,gerritsma12,mumford14a,mumford14b,joger14,ebgha19,chen2021}. In general this system is chaotic \cite{mumford14a,chen2021}, but in this paper we do not allow the qubit to exchange energy with the bosons (dispersive limit of the interaction) and this renders the model integrable. Our model is relevant to experiments with bosonic Josephson junctions, e.g.\ Bose-Einstein condensates (BECs) in double well potentials \cite{albiez05,Levy07,LeBlanc2011,trenkwalder16}, or spinor condensates with two internal states \cite{zibold10,Gerving12}, if an additional impurity atom or ion \cite{zipkes10} is added. Alternatively, the same Hamiltonian (Ising model with long-range interactions) can be realized with trapped ions, again with two internal states \cite{Bohnet16}, and again a distinguished `impurity' atom or ion should be added to the system. Solid state Josephson junctions might offer another route to realize the type of dynamics we will discuss here \cite{Zhu2001,DeFranceschi2010,Pal2018}. 

The structure of the rest of this paper is as follows: Sec.\ \ref{Sec:Model} presents the details of the model (bosons+probe) to be used and Sec.\ \ref{Sec:Results} examines how a TTC of operators in the subspace belonging to the probe can be written as a periodic Floquet operator acting purely on the bosons.  In Sec.\ \ref{Sec:classical} the classical dynamics generated by this Floquet operator is examined, including the  classical Lyapunov exponent which is a signature of chaos. Sec.\ \ref{Sec:quantum} turns to quantum properties:  Sec.\ \ref{SubSec:Eigenphases} examines the quasienergy level spacings of the Floquet operator and compares against the results of RMT, while Sec.\ \ref{SubSec:Survival} treats the quantum TTC as a survival amplitude which leads naturally to computations of the inverse participation ratio and further comparisons against RMT.  Conclusions are presented in Sec.\ \ref{Sec:Conclusion}. This paper also has three appendices where details of the calculations and some extra supporting results are given.

\section{\label{Sec:Model}Model}

Our model couples the two-mode Bose-Hubbard model, which describes $N$ interacting bosons hopping between two sites, to a single two-state atomic quantum dot (AQD), i.e.\ a qubit. In order to put the qubit and the bosons on a similar footing it is convenient to express the boson operators in terms of collective spin operators 
\begin{equation}
\hat{S}_\alpha \equiv 1/2 \sum_i^N \hat{\sigma}_\alpha^i
\end{equation}
where $\alpha \in \{x,y,z\}$.  Using the Schwinger representation the same collective spin operators can alternatively be defined via annihilation and creation operators acting on the sites:  If we label the two sites of the Bose-Hubbard model as $L$ (left) and $R$ (right) then the number difference between the left and right sites can be written $ ( \hat{b}^\dagger_L\hat{b}_L - \hat{b}^\dagger_R \hat{b}_R ) \equiv 2 \hat{S}_z$, where $\hat{b}_{L/R}^\dagger$ ($\hat{b}_{L/R}$) is the creation (annihilation) operator for a boson on the left/right site obeying the usual commutation relations $[\hat{b}_i,\hat{b}_j^\dagger] = \delta_{i,j}$ where $i,j \in [L,R]$. Similarly, we have $( \hat{b}_R^\dagger \hat{b}_L + \hat{b}^\dagger_L \hat{b}_R  ) \equiv  2 \hat{S}_x $ which takes a boson from one site and puts it onto the other (plus the reverse process to make the operator hermitian), thus producing mode coupling (tunneling). 
 
In this way one finds that, up to constant terms, the total Hamiltonian can be written (for full details see \cite{mulansky11}), 
\begin{equation}
\hat{H} = \hat{H}_B + \hat{H}_d + \hat{H}_{Bd}\;,
\label{eq:ham}
\end{equation}
where $H_B$, $H_d$ and $H_{Bd}$ are the $N$ boson, dot and boson-dot interaction Hamiltonians, respectively, and are given by 
\begin{eqnarray}
\hat{H}_B &=& k_z \hat{S}_z^2/(N+1) - \alpha_x \hat{S}_x + \alpha_z \hat{S}_z \label{eq:HB}\\
\hat{H}_d &=& -\Delta \left ( 1 + \hat{\sigma}_z \right )/2  \label{eq:Hd}\\
\hat{H}_{Bd} &=& \beta \hat{S}_x \left ( 1+\hat{\sigma}_z \right ) \label{eq:HBd}\, .
\end{eqnarray}
Here,  the AQD operators are single Pauli matrices and are distinguished from the boson Pauli matrices by the absence of a superscript. The parameters in this Hamiltonian have the following definitions: $k_z$ is the boson-boson interaction energy, $\alpha_x$ is the boson hopping/tunneling energy, $\alpha_z$ is the energy imbalance between the two boson modes, $\Delta$ is the energy imbalance between the two AQD modes and $\beta$ is the coupling energy between the AQD and the $N$ bosons. 

 It is important to note that there is no hopping term ($\hat{\sigma}_{x}$ operator) in the dot's Hamiltonian and so it cannot make transitions between its two states,  i.e.\ $[ \hat{\sigma}_z, \hat{H} ] = 0$, and this gives a dispersive interaction between the bosons and the AQD. The AQD therefore plays a somewhat passive role in the dynamics of the bosons: from $\hat{H}_{Bd}$ we see that the effect of the AQD is to modify the tunneling energy of the bosons, either having no effect or suppressing it depending on whether the AQD is initialized in its excited or ground state, respectively.  It has previously been shown that when the AQD is allowed to make transitions the classical dynamics displays chaos above a certain critical value of $\beta$ \cite{mumford14a}, and without these transitions the dynamics is regular.  Nevertheless, we will show in this paper that when AQD operators are used in Eq.\ \eqref{eq:TTC}, apparently chaotic dynamics emerge in the TTC due to the presence of the AQD.

\begin{figure}[t!]\centering
    \includegraphics[width=0.7\columnwidth]{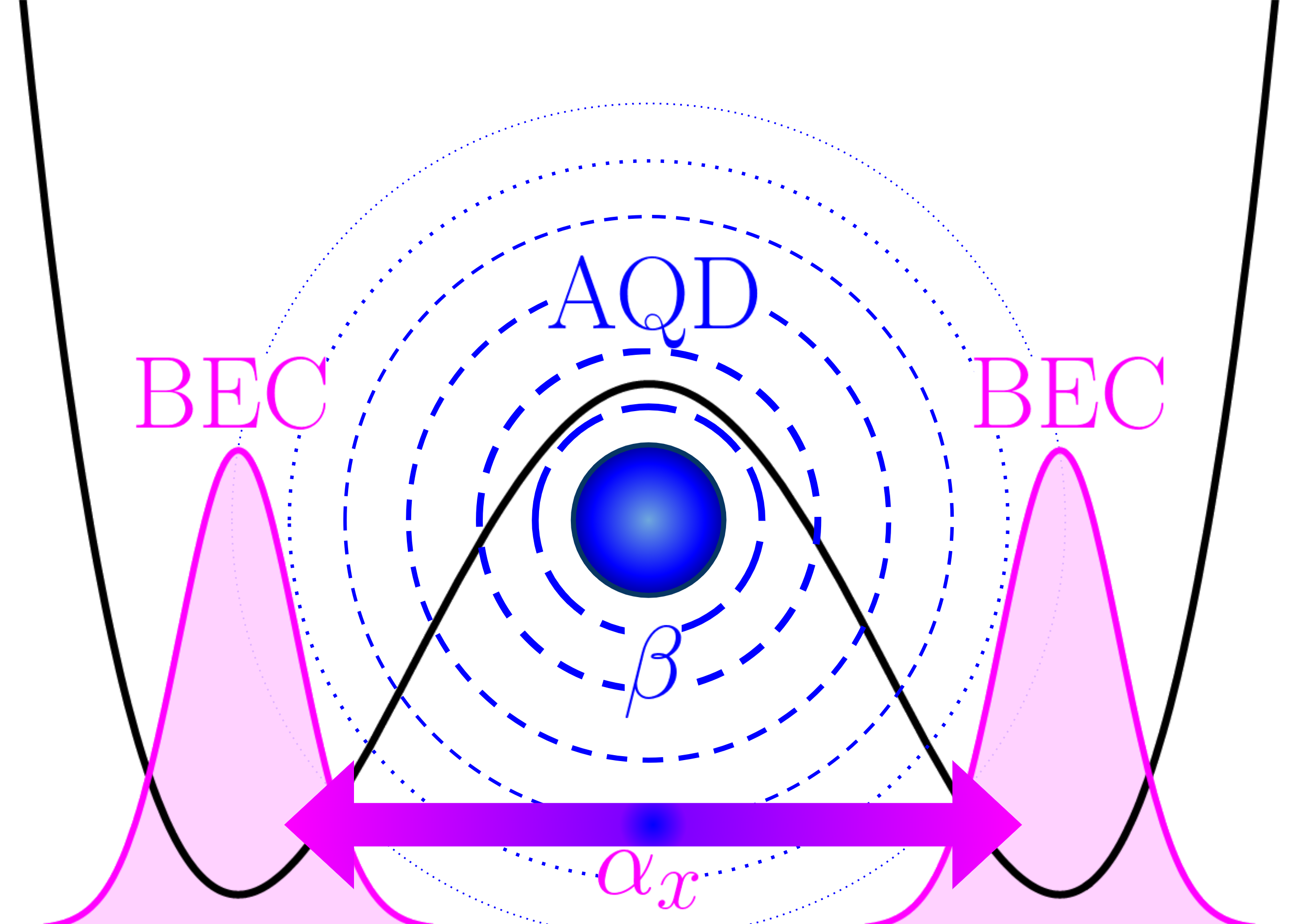}
	\caption{Schematic of system being considered: a BEC trapped in a double-well potential with inter-well hopping moderated by a trapped AQD. Tuning the parameter $\beta$ in $\hat{H}_{Bd}$ will effectively modify the hopping strength $\alpha_x$. The BEC has a self-interaction strength $k_z$, and a bias/tilt in the double-well can be controlled by $\alpha_z$.}
	\label{fig:Schematic}
\end{figure}

One way to realize the Hamiltonian in Eq.\ \eqref{eq:ham} is with a BEC trapped in a double well potential in the two-mode regime and coupled to a distinguishable atom trapped between the two wells, see Fig.\ \ref{fig:Schematic}.  This dot atom could be a different species to the rest of the atoms or simply be in a different hyperfine state. In fact the dot atom need not even be trapped between the two wells, and could instead be allowed to tunnel back and forth between the two wells like the bosons \cite{Rinck2011,mulansky11,chen2021}. However, in order for the interaction to be dispersive in this case the energy difference between the symmetric and antisymmetric states of the dot atom should be much greater than all the other energies so that it remains in just one state during the dynamics.

For such an ultracold atom realization the parameters in Eqs.\ \eqref{eq:HB}-\eqref{eq:HBd} can all be controlled using external fields: $k_z$ and $\beta$ contain the boson-boson and boson-dot s-wave scattering lengths, respectively, and can be controlled via the Feshbach resonance technique \cite{muessel15,perrin09}; $\alpha_x$ is the tunneling energy between the two BEC wells and can be controlled by raising or lowering the height of the barrier, e.g.\ by laser intensity; $\alpha_z$ is the energy imbalance between the two wells and can be controlled by providing a tilt between them via external fields with spatial gradients; $\Delta$ is the energy difference between the first two states of the AQD. An alternative realization, with slightly modified parameter definitions, makes use of internal states (for both the bosons and the AQD) rather than spatial states,  although all the atoms should then be tightly trapped so that they occupy a single spatial mode.  The tunneling between states in this case must be driven by laser or radio frequency radiation depending on whether the different internal states are different electronic orbitals or hyperfine states.

Yet another way to realize the boson-impurity model is via ions immersed in a BEC.  In the experiment discussed in Ref.\ \cite{zipkes10} a single ${}^{171}\mathrm{Yb}^{+}$ ion was trapped inside an  $^{87}\mathrm{Rb}$ BEC and the experiment demonstrated independent control of the ion and the BEC. Furthermore, temperatures where ion-atom collisions are dominated by the s-wave channel are being approached \cite{pinkas20}. The Hamiltonian $\hat{H}_B$ can also be realized using a linear ion trap system using ${}^{171}\mathrm{Yb}^{+}$ with effective magnetic fields generated by stimulated Raman transitions \cite{hess17}.  


\section{\label{Sec:Results}Correlation function dynamics: Floquet operator}


Our philosophy in this paper is to treat the AQD as a probe of the boson dynamics and hence the operators we shall use in Eq.\ \eqref{eq:TTC} are all AQD operators. In fact, for simplicity we use $\hat{\sigma}_x$ for both operators $\hat{A}$ and $\hat{B}$ where, of course, we evolve  $\hat{A}(t)=\hat{\sigma}_{x}(t)$ as a function of time but keep  $\hat{B}(0)=\hat{\sigma}_{x}(0)$ at $t=0$. 
Furthermore, we shall assume that at $t=0$ the probe and bosons are uncorrelated so that the initial state of the system is a product state of the form $\vert \Psi \rangle = \vert \psi \rangle_{B} \otimes \vert + \rangle_{d}$ where $\vert \psi \rangle_{B}$ is the state of the BEC and $\vert + \rangle_{d}$ is the excited state of the quantum dot (starting the dot in the ground state works as well). This is, therefore, the state we use for evaluating the correlator  $\langle \dots \rangle$. In addition, we make two simplifications to Eqs.\ \eqref{eq:HB}-\eqref{eq:HBd}: first, since $[\hat{H}_d ,\hat{H}] = 0$ the AQD Hamiltonian will only produce an overall dynamical phase in our calculations which will not affect the results, so we set $\Delta = 0$, and second, without loss of generality we set $\beta = \alpha_x/2$ (we shall explain this last condition below).

Under these conditions we find that the $n$-fold TTC reduces to a correlation function evaluated purely within the bosonic state
\begin{equation}
F_n(t) = {}_{B}\langle \psi \vert \hat{\mathcal{F}}^n  \vert \psi \rangle_{B} \;,
\label{eq:SOTOC}
\end{equation}
where $\hat{\mathcal{F}}^n$ is a Floquet operator that repeatedly applies ($n$ times) the unitary operator
\begin{equation}
\hat{\mathcal{F}} =  e^{-i \hat{H}_1 t} e^{-i \hat{H}_2 t} \; .  \label{eq:floquetoperator}
\end{equation}
The derivation of this result is given in Appendix \ref{app:TTCCalc} where it is shown that the
two Hamiltonians appearing in $\hat{\mathcal{F}}$ are given by
\begin{align}
\hat{H}_1 =\;& -\hat{H}_B = - k_z \hat{S}_z^2/(N+1) + \alpha_x \hat{S}_x - \alpha_z \hat{S}_z \label{eq:ham1} \\
\hat{H}_2 =\;& \hat{H}_B \big |_{\alpha_x = 0} =  k_z \hat{S}_z^2/(N+1)  + \alpha_z \hat{S}_z \label{eq:ham2}\; .
\end{align}
In line with the reduction of the full state to only the bosonic state in Eq.\ (\ref{eq:SOTOC}), it is notable that these effective Hamiltonians  depend only on the boson operators and are different versions of $\hat{H}_B$ given in Eq.\ (\ref{eq:HB}). Hence, the general TTC has turned into a survival amplitude for the BEC part $\vert \psi \rangle_{B}$   of the total state after $n$ applications of the operator $\hat{\mathcal{F}}$.  The different signs in front of the two Hamiltonians mean that the Floquet operator describes a system being shaken forwards and backwards (with slightly different forwards and backwards evolution) in time where the elapsed time $t = T$ is the length of each part of the shake and the order $n$ of the TTC is the number of cycles.

Our choice of $\beta=\alpha_{x}/2$ simplifies the Hamiltonian $\hat{H}_{2}$ so that $\hat{S}_{x}$ does not appear, as explained in Appendix \ref{app:TTCCalc}. This tuning is not central to the validity of our results: other choices will simply add a finite hopping term to $\hat{H}_{2}$. However, the vanishing of the $\beta$ term in the other Hamiltonian is a robust feature of having an uncorrelated initial state. This is a special choice, but also a very reasonable one. As long as the initial state is separable, one will always get forward and backward pieces like in Eqns. (\ref{eq:ham1}) and (\ref{eq:ham2}). 

We note in passing that a related issue to the one discussed in this paper is also an important consideration in digital quantum simulation, i.e.\ the breakup of continuous time evolution into separate steps. This sometimes goes under the name of Trotterization (after Trotter-Suzuki decompositions) and careful control of the errors induced by this process is critical to the accuracy of such simulations. In fact, it has been shown that there is a deep connection to chaos because chaotic systems are intrinsically less stable against such errors than nonchaotic ones \cite{sieberer19}.


%

%

\begin{figure}[t!]
	\includegraphics[width=0.49\columnwidth]{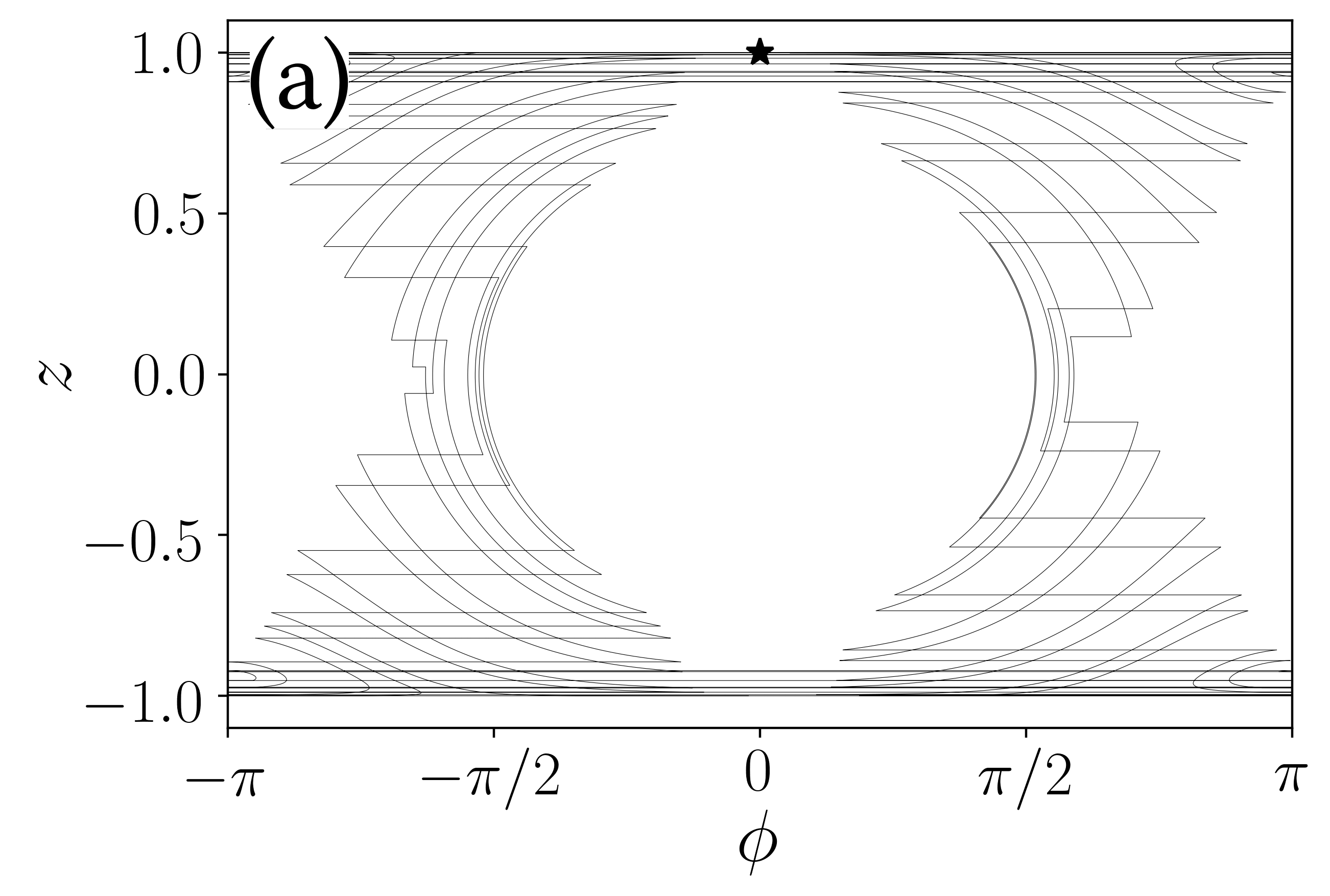}\includegraphics[width=0.49\columnwidth]{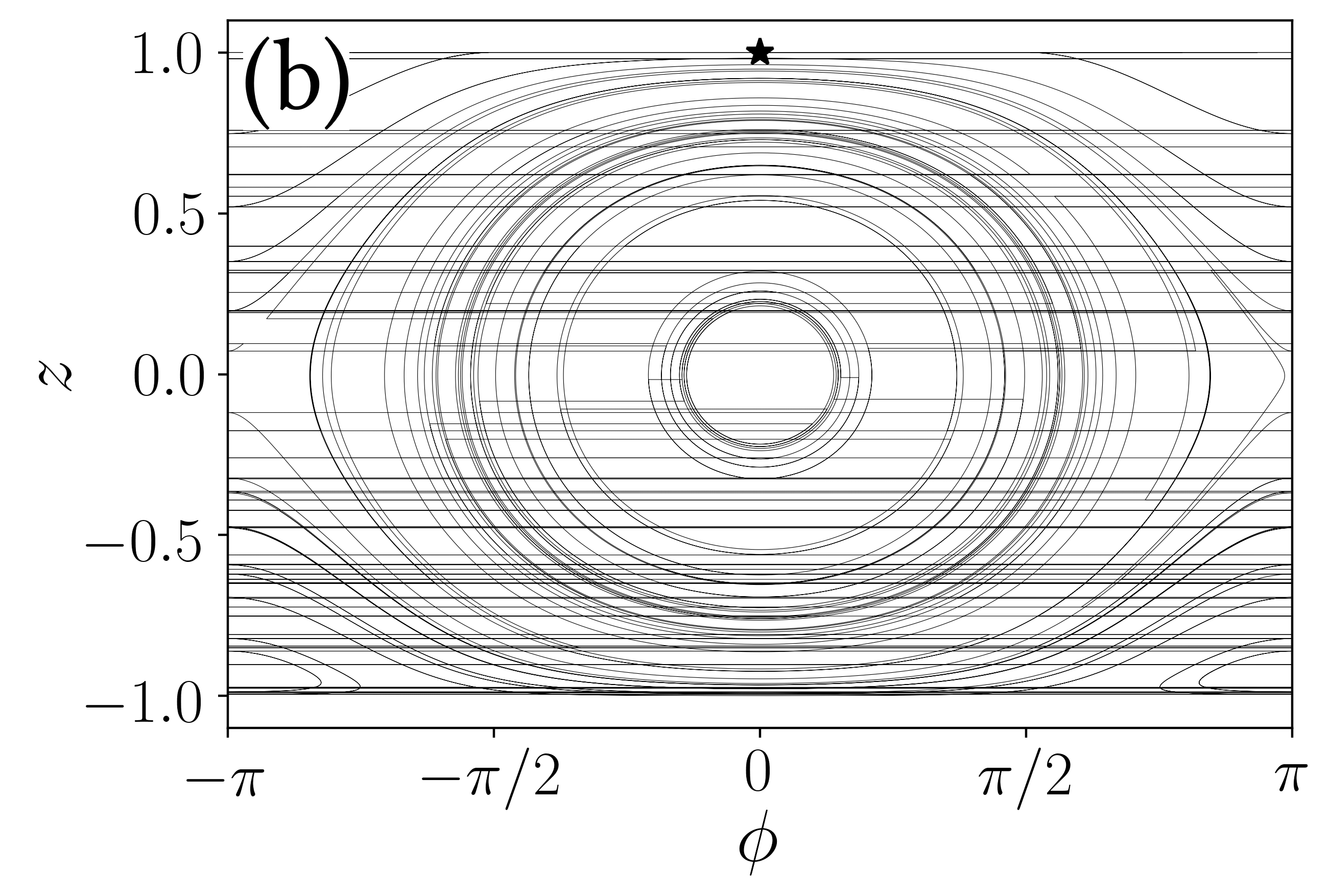}
	\includegraphics[width=0.49\columnwidth]{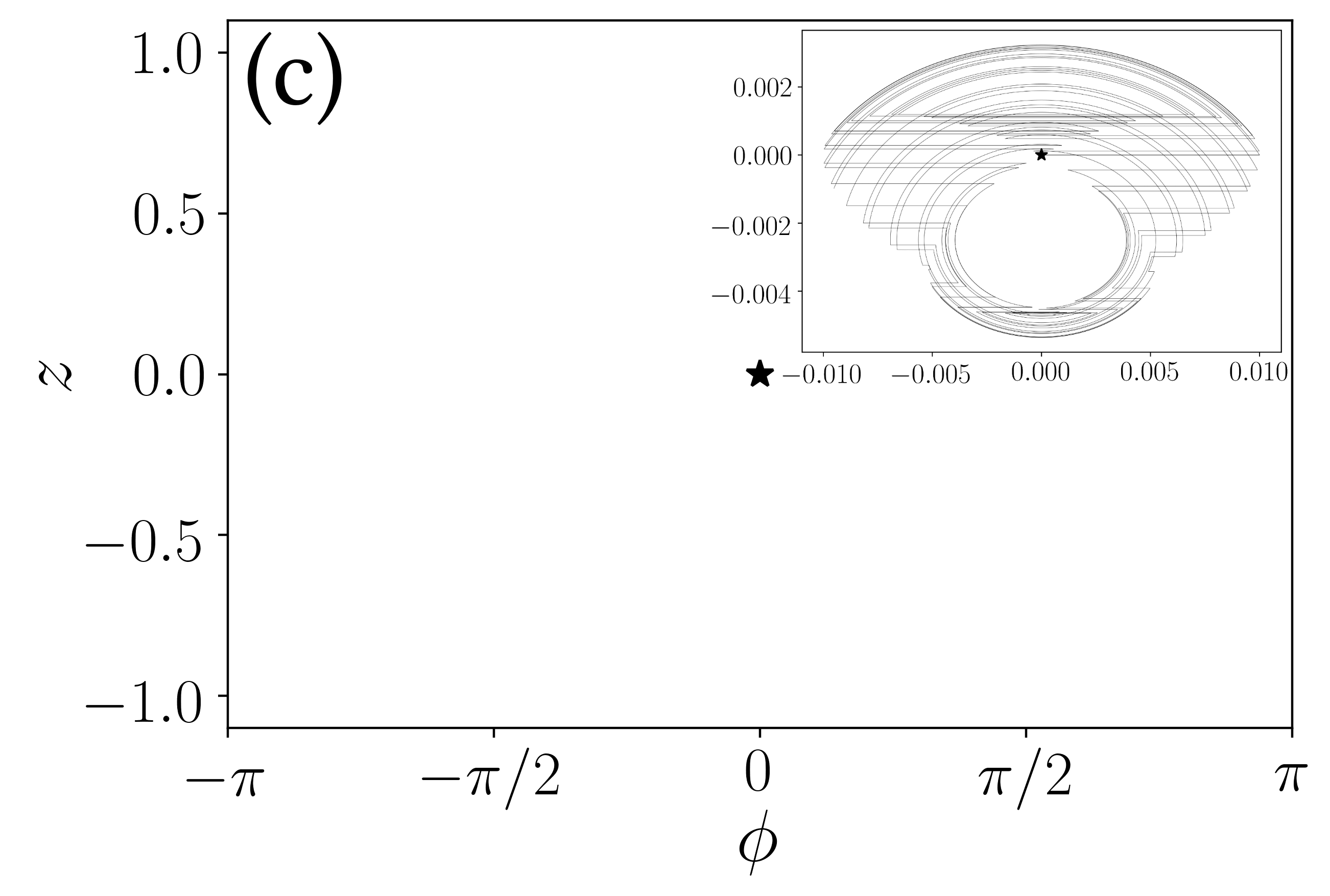}\includegraphics[width=0.49\columnwidth]{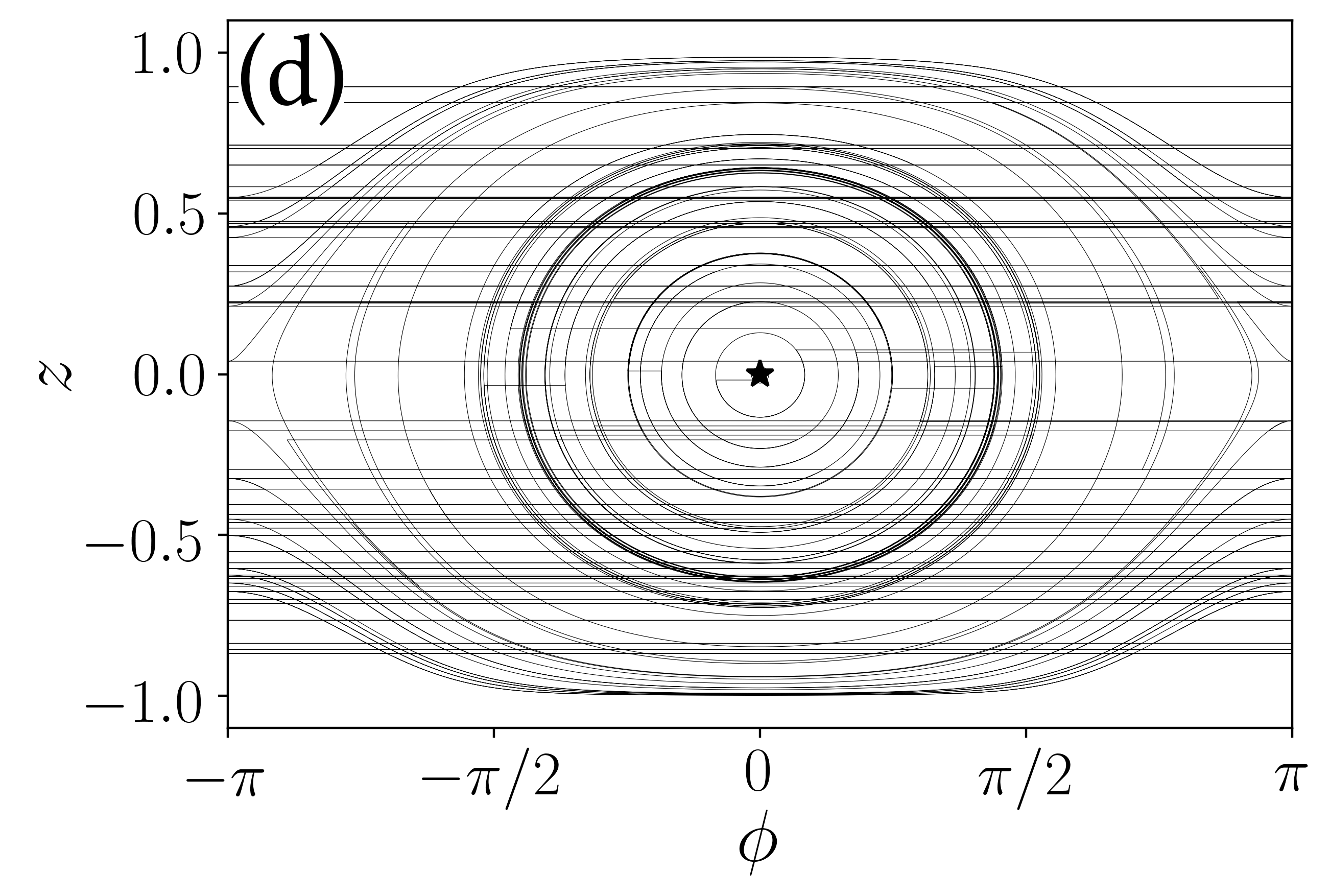}
	\includegraphics[width=0.49\columnwidth]{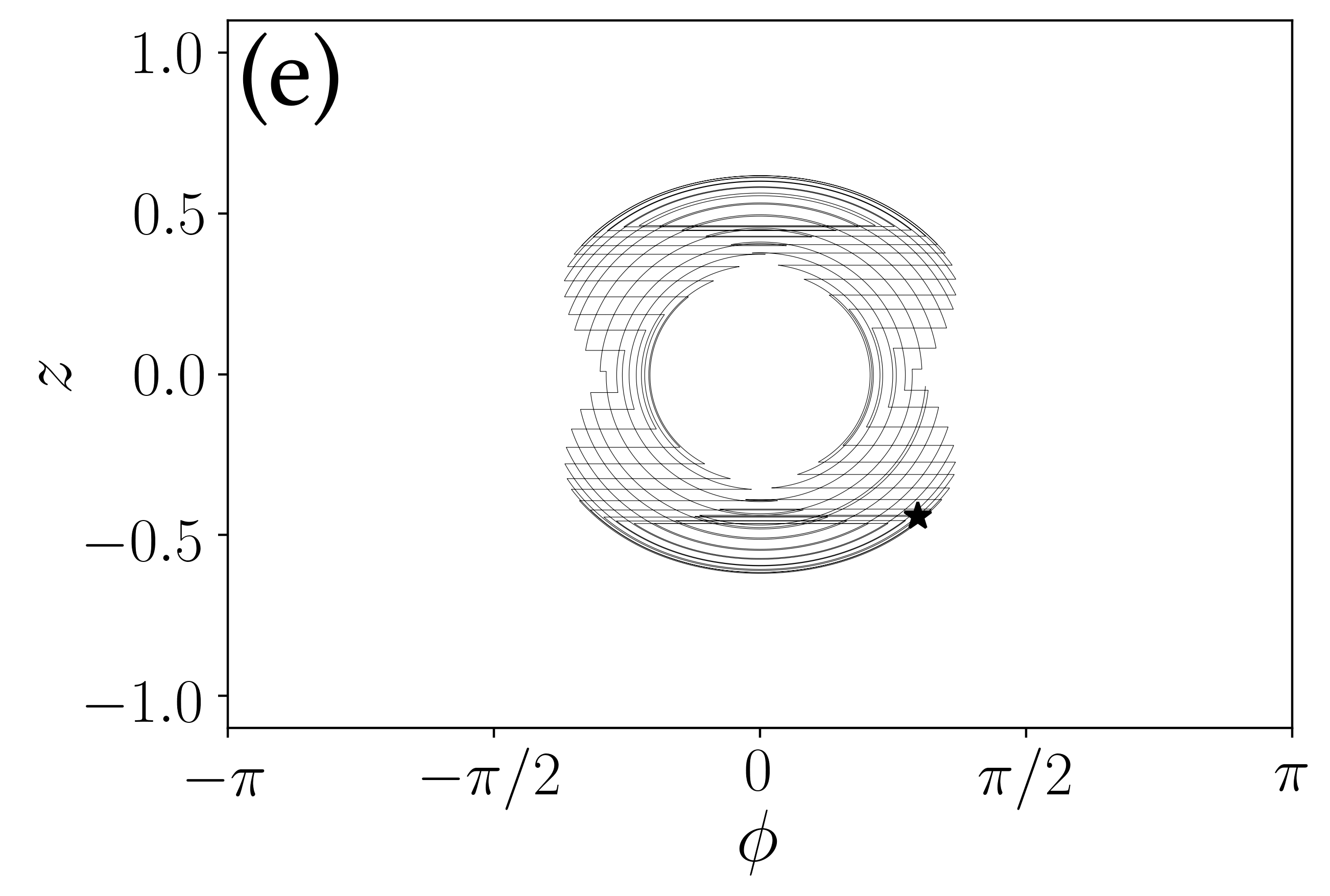}\includegraphics[width=0.49\columnwidth]{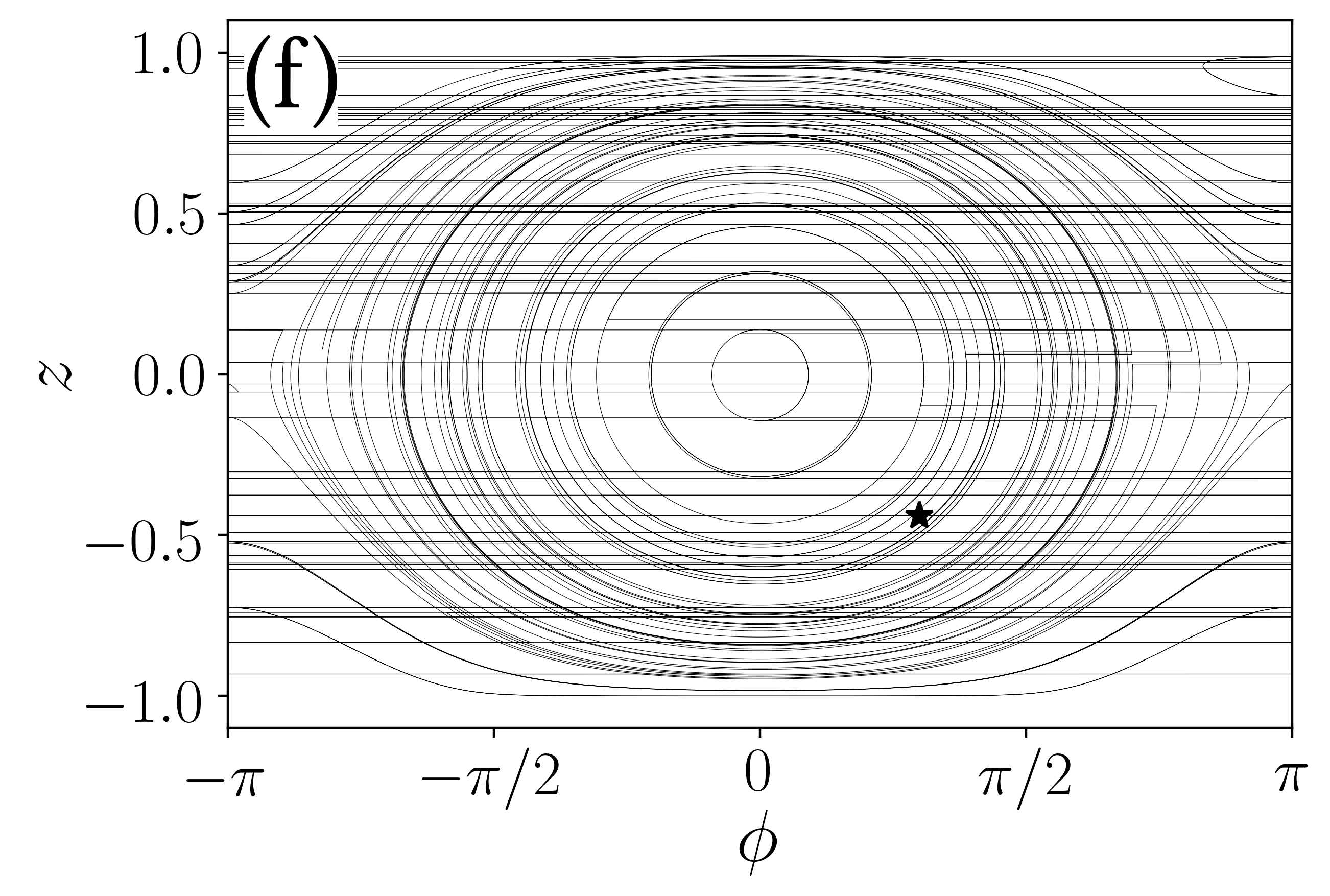}
	\caption{Classical phase space portraits of the $n$-fold TTC obtained using the mean-field Hamiltonians $\mathcal{H}_1$ and $\mathcal{H}_2$ given in Eqns.\ \eqref{eq:HMF1} and \eqref{eq:HMF2}, respectively. As explained in the text, the choices we make for the operators and states in the correlation function mean that the probe drops out of the problem and the correlation dynamics is purely for the bosons: $z$ is the number difference and $\phi$ the phase difference between the two bosonic modes. The star in each panel indicates the starting point. Time evolution proceeds with repetitive application of each Hamiltonian and since $\hat{H}_2$ is diagonal in the $\hat{S}_z$ basis, then in $z$ coordinates $\mathcal{H}_2$ will evolve the classical trajectory in a straight horizontal line to another constant energy contour of $\mathcal{H}_1$.  The left column shows classical trajectories with $\alpha_xT=1$ (weakly chaotic) while the right column shows $\alpha_xT=5$ (strongly chaotic). \textbf{Panels (a)-(b):} Initial state has all spins pointing along $S_z$ axis. \textbf{Panels (c)-(d):} Initial state has all spins pointing along $S_x$ axis. Panel (c) has such little variation from the initial state that it must be shown in an inset. \textbf{Panels (e)-(f):} Randomly selected initial classical vector with no special symmetry. For all panels, $k_z=3\alpha_x$, $\alpha_z=0.01\alpha_x$, and $n=50$.}
	\label{fig:PhaseSpace}
\end{figure}

\begin{figure}[t!]\centering
        \includegraphics[width=\columnwidth]{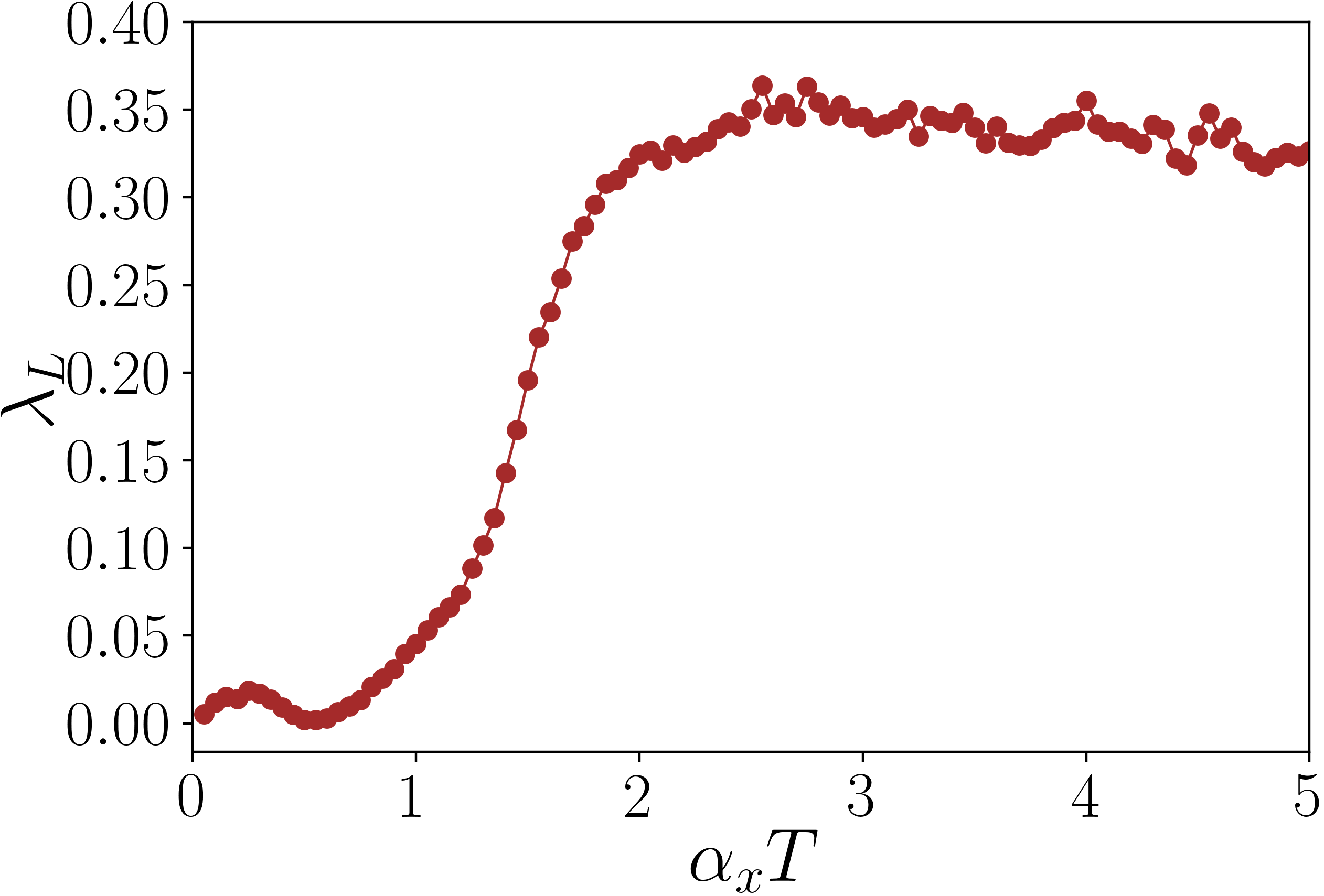}
        \caption{The Lyapunov exponent, $\lambda_L$, as a function of the classical kick period $\alpha_xT$ for dynamics produced by the repeated application of the mean-field Hamiltonians in Eqns.\ \eqref{eq:ham1} and \eqref{eq:ham2}. At short times, $\lambda_L \approx 0$ suggesting the system is regular while for longer times a clear nonzero Lyapunov exponent develops, suggesting chaotic dynamics.  Each data point is the maximum $\lambda_L$ which is then averaged over 1500 random initial states in phase space. Here, $\alpha_z = 0.01 \alpha_x$ and $k_z = 3 \alpha_x$ while the Hamiltonians $\mathcal{H}_1$ and $\mathcal{H}_2$ are cycled $n = 20$ times. These results can be contrasted with those shown in Fig.\ \ref{fig:LEvsk} in Appendix \ref{app:MeanField} where it is shown that the true (continuous) dynamics of the full Hamiltonian has $\lambda_L=0$.}
\label{fig:LEPic}
\end{figure}

\section{\label{Sec:classical}Classical correlation function dynamics}

We begin our analysis of the correlation function dynamics by checking for classical chaos. The classical theory is given by the mean-field approximation and it is known that in this limit the two-mode Bose Hubbard model, which describes the bosonic part of the system, is equivalent to a non-rigid pendulum which is an integrable system \cite{Smerzi97}. When coupled to an impurity spin the total system can be mapped in certain regimes onto a double pendulum \cite{mulansky11} which is in general chaotic, although in the present dispersive case the second pendulum has a constant angular momentum which keeps the model integrable.

The mean-field versions of our various Hamiltonians are derived in Appendix \ref{app:MeanField}. For example, it is shown that the mean-field version $\mathcal{H}$ of the total  Hamiltonian $\hat{H}$ is given by Eq.\ (\ref{eq:MFHamil}) which is obtained by replacing the mode operators with complex numbers and taking the thermodynamic limit $\mathcal{H} = \lim_{N \to \infty} \hat{H}/N$.  From there, Hamilton's equations of motion can be used to generate the dynamics.  In the case of $\mathcal{H}_1$ and $\mathcal{H}_2$ (the mean-field limits of $\hat{H}_1$ and $\hat{H}_2$, respectively), the dynamics are produced by using Hamilton's equations of motion for $\mathcal{H}_2$ for a period of time $T$, then we switch to Hamilton's equations for $\mathcal{H}_1$ for the same period of time, and repeat this intertwining process for $n$ cycles. 

In Fig.\ \ref{fig:PhaseSpace} we plot the phase space dynamics of the BEC mean-field variables 
\begin{equation}
z = (n_L - n_R)/N \, , \quad \phi = \phi_L - \phi_R
\end{equation}
 which are the scaled number difference and phase difference between the left and right wells, respectively.  Each row has a different initial condition represented by a black star.  At short times (left column) the dynamics only accesses limited regions of phase space, especially if the system is initialized near the stable fixed point at $\phi=z=0$. However, at longer times (right column) the dynamics becomes ergodic and independent of the initial conditions which is a hallmark of classical chaos.  In fact, Fig.\ \ref{fig:PhaseSpace}  shows how ergodicity is established despite the two parts of the time evolution being separately integrable:  the $\mathcal{H}_2$ trajectories conserve $z$, so they travel only along the $\phi$ direction, essentially providing a pathway to different energy contours of $\mathcal{H}_1$, along which the orbit proceeds.

To quantify the chaotic dynamics we numerically compute the Lyapunov exponent $\lambda_L$ and the results are plotted in Fig.\ \ref{fig:LEPic} as a function of $\alpha_x T$. Each data point is obtained from  a phase space average taken with respect to many random initial states. When $\lambda_L > 0$ the trajectories are exponentially separating in time and hence we have chaos.  As expected from Fig.\  \ref{fig:PhaseSpace}, for relatively short periods ($\alpha_x T \lesssim 1$) the trajectories tend to stay in a small subregion of phase space leading to $\lambda_L \approx 0$. As the shake period is increased ($\alpha_xT\gtrsim 1$) the Lyapunov exponent becomes nonzero and the correlator dynamics  becomes chaotic.  For a comparison with the actual model see Fig.\ \ref{fig:LEvsk} in Appendix \ref{app:MeanField} where a plot of $\lambda_L$ versus $k_z$ shows that $\lambda_L=0$, demonstrating that $\mathcal{H}$ is regular. These results give us the first hint that the dynamics of the general TTC function can be chaotic even for a nonchaotic system.

\section{\label{Sec:quantum}Quantum correlation function dynamics}

We now turn to the fully quantum problem to look for evidence of so-called `quantum chaos' in the correlator dynamics. To this end we first examine the spectral statistics of the Floquet operator and compare to RMT and second we calculate the survival probability. 

\subsection{Eigenphases and Spacings of the Floquet Operator}
\label{SubSec:Eigenphases}

We first consider the Floquet operator $\hat{\mathcal{F}}$ in Eq.\ \eqref{eq:floquetoperator}.  Recently it was shown that in a quantum stadium billiard model (which has a classical limit which is chaotic) the spectral statistics of the operator $\hat{\Lambda}(t) = \mathrm{ln}\left ( -[\hat{x}(t),\hat{p}_x(0)]^2 \right )/(2t)$, which contains an out-of-time-ordered \textit{commutator}, align well with the predictions from RMT \cite{rozenbaum19}.  Similar results were found when analyzing the spectral statistics of a Floquet operator for a shaken system when its corresponding classical system is chaotic \cite{sieberer19}.  We will follow this route here and analyze the spectrum of $\hat{\mathcal{F}}$. This is a unitary operator, but rather than work with its complex eigenvalues $e^{i\theta_j }$, we instead examine the statistical properties of its eigenphases $\theta_j$.  

\begin{figure}
	\includegraphics[width=0.9\columnwidth]{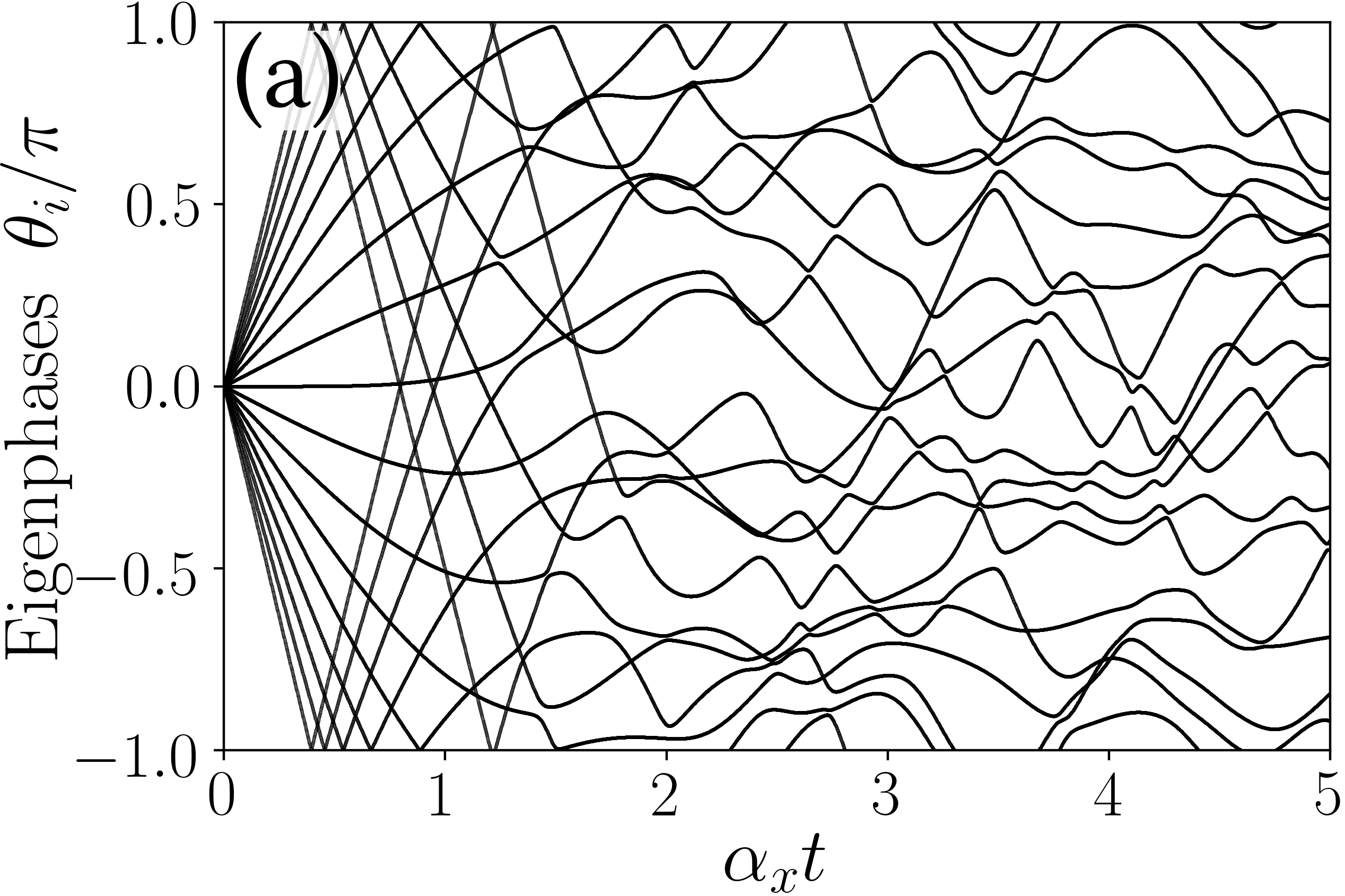}
	\includegraphics[width=0.9\columnwidth]{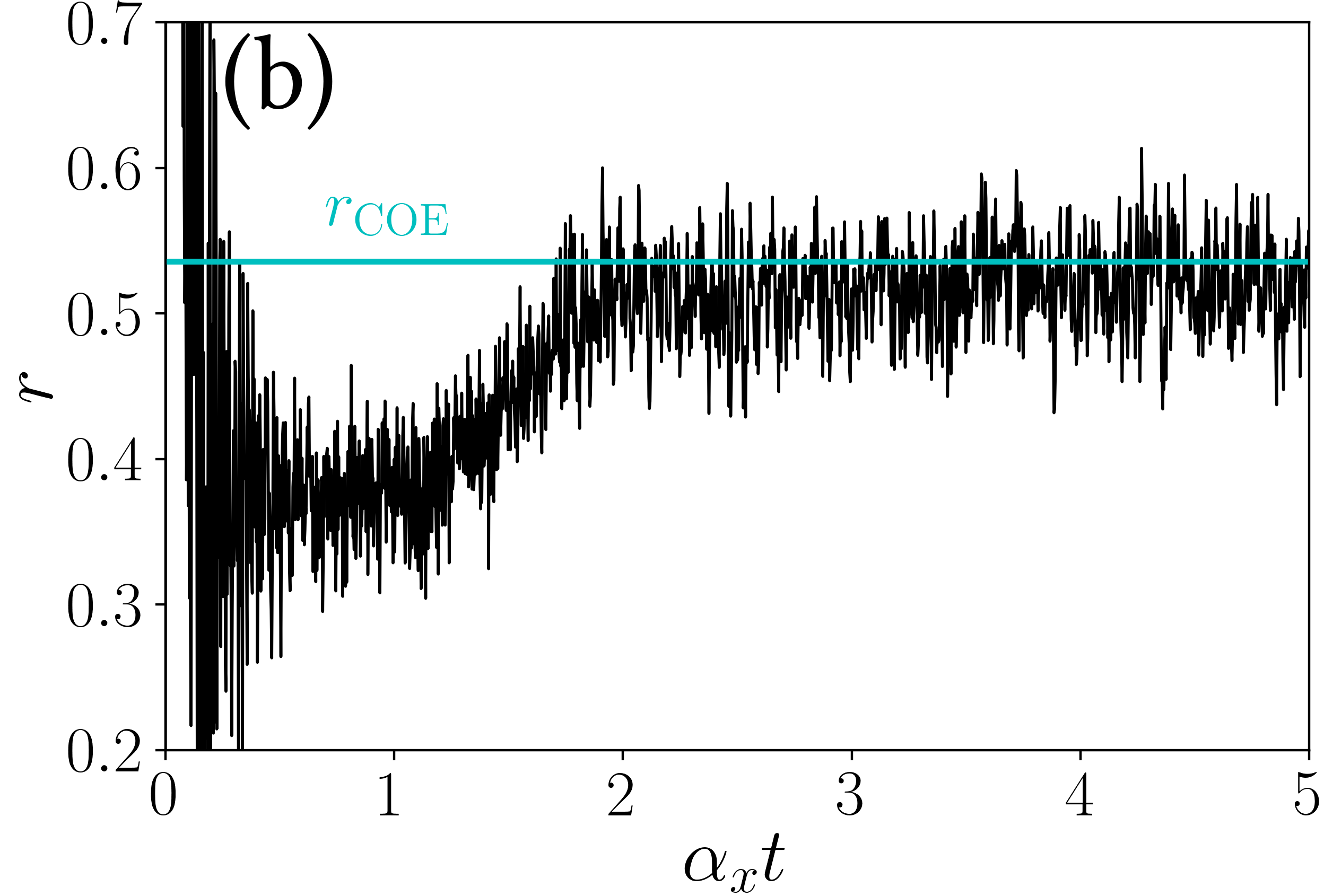}
	\caption{\textbf{Panel (a):} Eigenphases $\theta_i$ of the Floquet operator $\hat{\mathcal{F}}$ as a function of time for $N = 16$.  At early times the eigenphases evolve regularly, but as time goes on they begin to wind around the interval $[-\pi, \pi)$ and avoided crossings form. These are initially tiny but gradually widen and result in the familiar level repulsion seen in chaotic systems. \textbf{Panel (b):} Average eigenphase spacing ratio $r$ as a function of time for $N = 100$.  At later times $r$ oscillates around the RMT prediction of $r_\mathrm{COE}$, shown as a horizontal cyan line, coinciding with the occurence of level repulsion in (a).  The other parameters for both images are $k_z = 3 \alpha_x$ and $\alpha_z = 0.01 \alpha_x$. }
\label{fig:EPrr}
\end{figure}

First, we note that the eigenphases are time dependent in a nontrivial way which can be seen by writing $\hat{\mathcal{F}}$ in terms of a single effective Hamiltonian $\hat{\mathcal{F}} = e^{-i\hat{H}_\mathrm{eff} t} = e^{-i\hat{H}_1 t} e^{-i\hat{H}_2 t}$.  Using the Baker-Campbell-Hausdorff formula,
\begin{equation}
e^{\hat{X}} e^{\hat{Y}} =  \mathrm{exp} \left ( \hat{X} + \hat{Y} + \frac{1}{2} [\hat{X}, \hat{Y}] + \dots \right )
\end{equation}
we find that $\hat{H}_\mathrm{eff}$ for our system can be written at short times as
\begin{equation}
\hat{H}_\mathrm{eff}(t) =   \hat{H}_1 + \hat{H}_2  -\frac{it}{2} [\hat{H}_1, \hat{H}_2] + \mathcal{O}(t^2) \, .
\label{eq:heff}
\end{equation}
At each moment in time the effective Hamiltonian yields a set of instantaneous eigenstates $\{ \vert v_i (t) \rangle \}$ such that $\hat{H}_\mathrm{eff} (t) \vert v_i (t)\rangle = \epsilon_i (t) \vert v_i (t) \rangle$ where $\{\epsilon_i(t)\}$ is the set of instantaneous eigenenergies.  From Eq.\ \eqref{eq:heff}, we can see at early times the dynamics is simply due to $\exp[-i(\hat{H}_1 + \hat{H}_2 ) t]$, however, as $t$ increases more terms contribute to $\hat{H}_\mathrm{eff}$ and the dynamics of the eigenenergies become complicated.  Since $\epsilon_i t = \theta_i  \pmod{2\pi}$, we expect the eigenphases to inherit this complicated behavior and this can indeed be seen in Fig.\ \ref{fig:EPrr}(a) where they are plotted as a function of time.  At early times the magnitude of the eigenphases increases linearly until at around $\alpha_x t \approx 0.4$ they begin to wind around the interval $[-\pi, \pi)$ at which point tiny avoided crossings form (they are at first too small to see at the scale of the the figure, however when zoomed-in incredibly narrow avoided crossings become visible).  At around $\alpha_x t \approx 1.3$ the avoided crossings begin to widen and at late times the eigenphases are well separated and display the equivalent of level repulsion found in time-independent chaotic systems.

Rather than examine the full statistical distribution of the eigenphases, we instead calculate the average spacing ratio \cite{oganesyan07}, 
\begin{equation}
r_n = \frac{\mathrm{min}(\delta_n, \delta_{n+1})}{\mathrm{max}(\delta_n, \delta_{n+1})}, \hspace{20pt} r = \frac{1}{\mathcal{D}} \sum_{n = 1}^\mathcal{D} r_n
\end{equation}
where $\delta_n = \theta_{n+1} - \theta_n$ is the difference between adjacent eigenphases and $\mathcal{D} = N + 1$ is the size of the BEC Hilbert space.  The spacing ratio takes on distinct values depending on which RMT ensemble the eigenphases follow, if they follow any at all.  For our case, having $\alpha_z \neq 0$ destroys the parity symmetry of the system leaving only time reversal symmetry. We therefore expect that $r$ should obey the circular orthogonal example (COE) result of $r_\mathrm{COE} = 4-2 \sqrt{3} \approx 0.536$.  Fig.\ \ref{fig:EPrr}(b) shows clearly that $r$ does indeed oscillate around the COE result provided we consider longer times $\alpha_x t > 2.0$.  In fact, comparing panels (a) and (b) in Fig.\ \ref{fig:EPrr} we see that the dip in $r$ in the range $0.4 \leq \alpha_x t \leq 1.3$ corresponds to the range of times when the eigenphases first begin to wind around the full interval $[-\pi, \pi)$ and form small avoided crossings. The time at which the dip occurs and how low it is are nonuniversal features that depend on the parameters of the system.  For $\alpha_x t > 1.3$ the avoided crossings of the eigenphases begin to widen and eventually show the chaotic result of level repulsion quantified by $r \approx r_\mathrm{COE}$.

\subsection{Survival Probability}
\label{SubSec:Survival}

Returning to Eq.\ \eqref{eq:SOTOC}, we focus our attention back on $F_n(t)$. More precisely, we study its squared absolute value which corresponds to the return probability $P_n(t) = \vert F_n(t) \vert^2$.  Furthermore, based on the results of the last section we expect the clearest evidence of chaos to come from the long time average of $P_n(t)$.  We start by inserting the resolution of identity,  $\mathbbm{1} = \sum_i \vert v_i\rangle\langle v_i\vert$ (expressed in terms of the eigenvectors of $\hat{H}_{\mathrm{eff}}$),  into $P_n(t)$. This gives
\begin{eqnarray}
	P_n(t) &=& \Big \vert \sum_{i = 1}^\mathcal{D}  \big \langle \psi \vert \hat{\mathcal{F}}^{n} \vert v_i \rangle \langle v_i  \vert \psi \rangle  \Big \vert^2 \nonumber \\
	&=&   \sum_{i, j}^\mathcal{D} e^{i n (\epsilon_i - \epsilon_{j})t}  \vert \langle v_i  \vert \psi \rangle \vert^2  \vert \langle v_{j}  \vert \psi \rangle \vert^2  \, .
\label{eq:SP1}
\end{eqnarray}
Here, and from now on, we suppress the subscript ``B'' on $\vert \psi \rangle_{B}$.
Since the eigenstates and eigenenergies of $H_\mathrm{eff}$ are complicated functions of time, even the qualitative behaviour of the long time average of Eq.\ \eqref{eq:SP1} is not immediately clear.  For large enough TTC order $n$, the phase factor will oscillate rapidly making all terms where $\epsilon_i \neq \epsilon_j$ approximately equal to zero (the ``diagonal approximation'').  This condition is satisfied naturally in the chaotic regime due to level repulsion of the eigenenergies. The level repulsion of the corresponding eigenphases is shown in Fig.\ \ref{fig:EPrr}, so we should expect the same for the eigenergies resulting in no degeneracies in the spectrum.  Thus, when $n$ is large enough, the long time average of $P_n(t)$, $P = \lim_{T \to \infty} \frac{1}{T} \int_0^T P_n(t) dt$, can be written as
\begin{eqnarray}
	P &\approx&  \sum_{i, j}^\mathcal{D} \delta_{i, j} \vert \langle v_i \vert \psi \rangle \vert^2 \vert \langle \psi \vert v_j \rangle \vert^2 \nonumber \\
	&=& \mathrm{IPR}\{\vert \psi \rangle\} \, .
	\label{eq:SP}
\end{eqnarray}
Thus, the survival probability becomes equal to the inverse participation ratio (IPR) of the state $\vert \psi \rangle$ over the basis states of $\hat{H}_\mathrm{eff}$ (or $\hat{\mathcal{F}}$) where the participation ratio (PR) is defined as
\begin{equation}
\mathrm{PR}\{\vert \psi \rangle \} \equiv \frac{1}{\sum_a^\mathcal{D} \vert \langle v_a  \vert \psi \rangle \vert^4}\;,
\end{equation}
and is used to quantify how spread a state of interest $\vert \psi \rangle$ is over a reference basis $\{ \vert v_a \rangle \}$. What remains to be done is to explore the effect of different BEC states $\vert \psi \rangle$ in which the correlation function is evaluated, and we shall see that this choice can affect the outcome of $P$.

A generic state $\vert \psi \rangle$ (e.g.\ one taken at random, absent any special symmetry), has complex coefficients in the basis of $\hat{S}_z$ (i.e.\ the set of Fock states $\{\ket{m}\}$ where the eigenvalues $\{ m \}$ are half  the boson number difference between the two sites) and is best modelled by CUE, whose states are uniformly distributed on the unit sphere in $\mathbb{C}^\mathcal{D}$. If, however, we select a single eigenstate $\ket{m}$ of the $\hat{S}_z$ operator for our TTC, there is a shift instead to COE statistics due to the fact that these are eigenstates of $\hat{H}_2$, ultimately changing the symmetry of $\hat{\mathcal{F}}$. This can be seen explicitly from the TTC, using Eq.\ \eqref{eq:SOTOC},
\begin{align}
	 & F_n(t) =\; \bra{m} \big( e^{-i \hat{H}_1 t} e^{-i \hat{H}_2 t} \big)^{n} \ket{m} \nonumber \\
	&=\; e^{-i \phi(m) t} \bra{m} \underbrace{  \big( e^{-i \hat{H}_1 t} e^{-i \hat{H}_2 t} \big)^{n-1} e^{-i \hat{H}_1 t}}_{\hat{U}_t}\ket{m}\;,
	\label{eq:SOTOC2}
\end{align}
where the phase $\phi(m) =   k_z m^2/(N+1)  + \alpha_z m$ can be neglected because we are interested in the survival probability $P_n(t) = \vert F_n(t) \vert^2$.  The evolution operator $\hat{U}_t$ in Eq.\ \eqref{eq:SOTOC2} is related to the Floquet operator by
\begin{equation}
	\hat{U}_t = \hat{\mathcal{F}}^{n} e^{i \hat{H}_2 t}\;,
\end{equation}
and has the additional symmetry $\hat{U}_t = \hat{U}_t^T$, meaning its eigenstates have real components. Therefore, when considering the $\hat{S}_z$ eigenstates as our basis, the time-dependent eigenstates of $\hat{U}_t$ are not taken from matrices in the CUE (since those states have complex components) and instead are best modelled by random matrices in the COE, for which the states are distributed uniformly on the unit sphere in $\mathbb{R}^\mathcal{D}$.

\begin{figure}[t!]
	\includegraphics[width=\columnwidth]{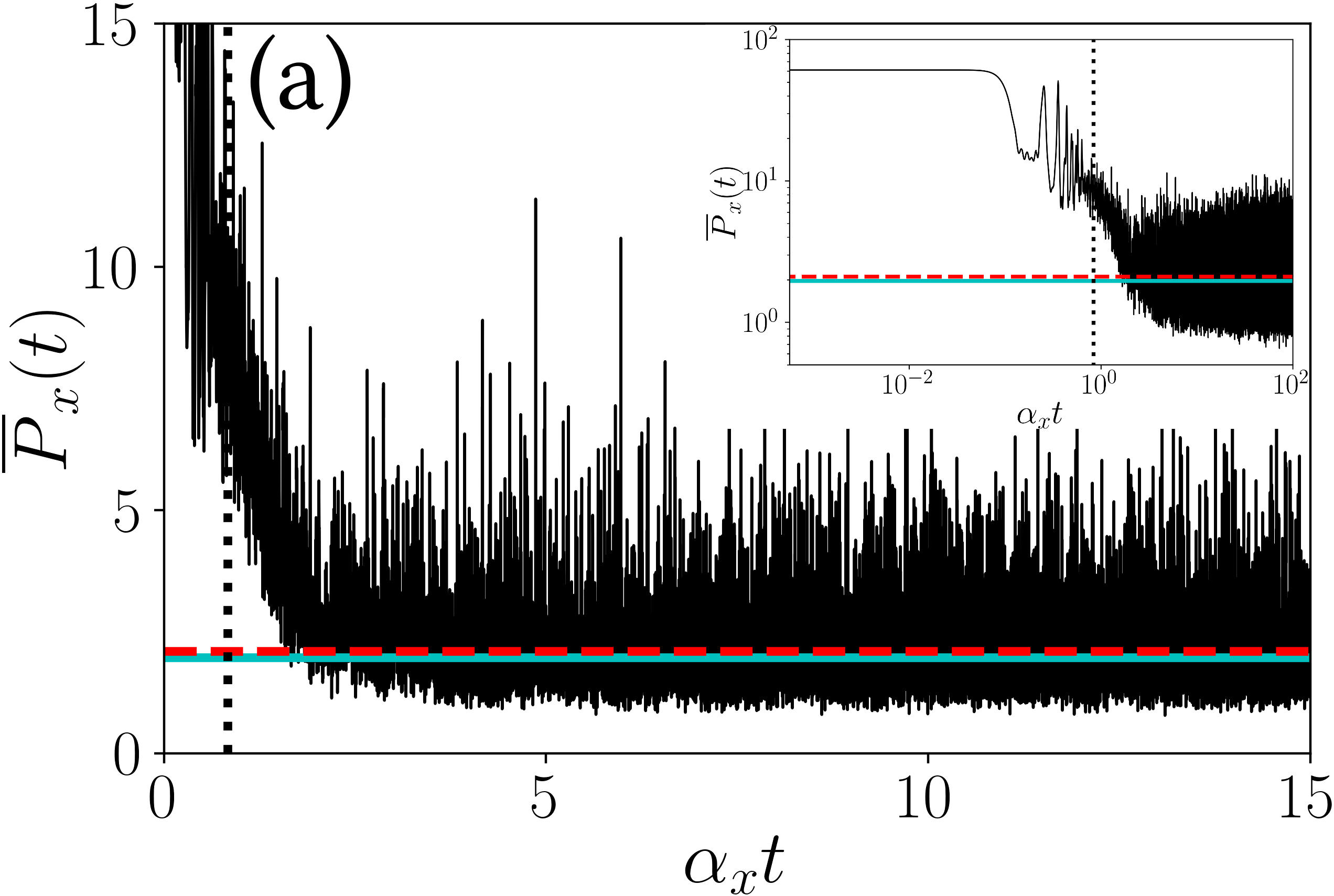}
	\includegraphics[width=\columnwidth]{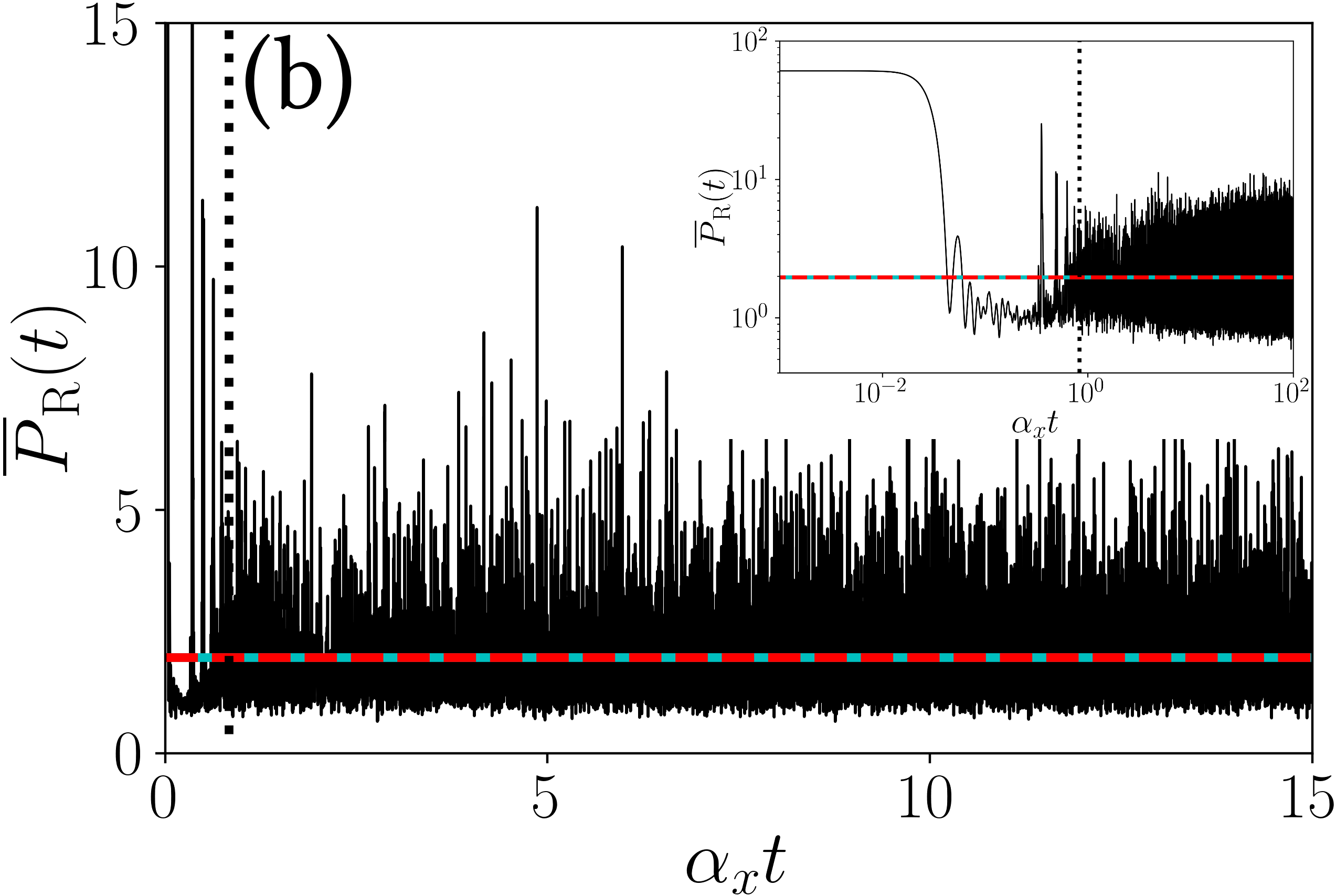}
	\includegraphics[width=\columnwidth]{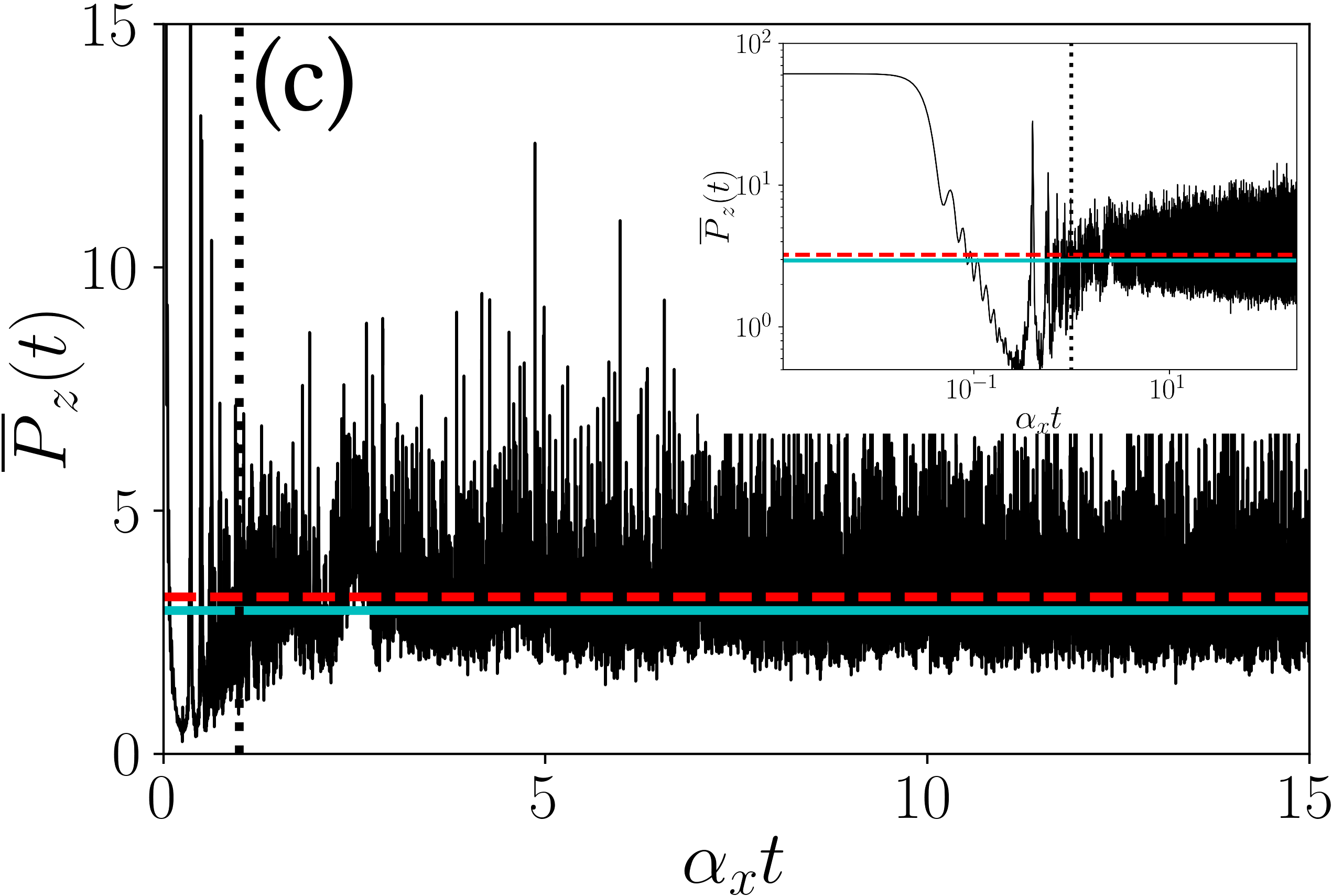}
	\caption{The survival probability averaged over bases as a function of time. \textbf{Panel (a):} Averaged over the $\hat{S}_x$ basis. The long-time average $\overline{P}_x\approx 2.106$ is shown as a red-dashed line, while the predicted $\mathrm{IPR}_{\mathrm{CUE}}\approx 1.968$ is shown as a solid cyan line. \textbf{Panel (b):} Averaged over a randomly selected basis. Here, the long-time average is $\overline{P}_{\mathrm{R}}\approx 1.971$. \textbf{Panel (c):} Averaged over the $\hat{S}_z$ basis. Here, $\overline{P}_z \approx 3.242$ and $\mathrm{IPR}_{\mathrm{COE}} \approx 2.905$, are shown as red dashed lines and cyan solid lines, respectively. Insets are identical, but plotted on a log-log scale, highlighting the transition from regular to chaotic dynamics. The Thouless time $t_{\mathrm{Th}}$ is shown in every panel as a vertical dotted line. The parameter values in all images are $k_z = 3 \alpha_x$, $\alpha_z = 0.01 \alpha_x$, $N = 60$ and $n = 50$.}
	\label{fig:PxPz}
\end{figure}

There exist signatures of chaos in both the survival probability of a single state, $P_n$, and for ensembles of states. For the purposes of this article, we will focus on general features of the survival probability and so we will average over an \textit{entire} basis, however a more in depth discussion of individual survival probabilities can be found in Appendix \ref{app:SurvProb}. Performing the average over a basis $\{\ket{\psi_i}\}$, we can write the sum over individual $\text{IPR}\{\ket{\psi_i}\}$ as
\begin{eqnarray}
\mathrm{IPR} = \sum_i \mathrm{IPR}\{\vert\psi_i \rangle \} = \sum_{i, j} \vert \langle v_j \vert \psi_i \rangle \vert^4 = \overline{P} 
\label{eq:AIPR}
\end{eqnarray}
where the bar over $P$ signals the average over the entire basis. If we use a random basis or the basis states of the $\hat{S}_x$ operator \cite{sieberer19}, we expect the average IPR to take on the CUE prediction,
\begin{equation}
	\mathrm{IPR}_{\mathrm{CUE}} = \frac{2 \mathcal{D}}{\mathcal{D} +1}
	\label{eq:IPRCUE}
\end{equation}
where in the thermodynamic limit, $N \to \infty$, and hence $\mathrm{IPR}_{\mathrm{CUE}} \to 2$. Meanwhile, if we use the basis of $\hat{S}_z$ eigenstates, we expect the average IPR to take on the COE prediction,
\begin{equation}
	\mathrm{IPR}_{\mathrm{COE}} = \frac{3\mathcal{D}}{\mathcal{D}+2},
	\label{eq:IPRCOE}
\end{equation}
for which $\mathrm{IPR}_{\mathrm{COE}} \to 3$ as $N \to \infty$.

Following Schiulaz \textit{et al.} \cite{schiulaz19}, we can also calculate the Thouless time $t_{\mathrm{Th}}$ for a fully connected model like ours in the COE and CUE ensembles. $t_{\mathrm{Th}}$ describes the time at which the wavefunction first extends over the entire many-body Hilbert space, and is thus the time at which universal RMT predictions begin to govern the dynamics. For spatially extended systems, $t_{\mathrm{Th}}$ generally depends on the system size \cite{schiulaz19,sierant20}, however for a fully connected model 
\begin{align}
	t_{\mathrm{Th}}^{\mathrm{CUE}}=&\;\left(\frac{3}{2\pi}\right)^\frac{1}{4}\label{eq:ThouCUE}\\
	t_{\mathrm{Th}}^{\mathrm{COE}}=&\;\left(\frac{3}{\pi}\right)^\frac{1}{4}\label{eq:ThouCOE}\;,
\end{align}
where these times are expressed in units of $\alpha_{x}^{-1}$.

In Fig.\ \ref{fig:PxPz}(a) we plot numerical evaluations of $P_n(t)$ averaged over the $\hat{S}_x$ basis. Since at short times $H_{\text{eff}}\approx \hat{S}_x$, the survival probability is briefly constant, and subsequently drops off to fluctuate near its relaxation value. The red dashed line is the long-time average and takes the value $\overline{P}_x \approx 2.105$, an approximately $7\%$ error from its expected value of $\mathrm{IPR}_{\mathrm{CUE}}|_{N=60}\approx 1.968$, shown as a solid cyan line.  Although the survival probability has large amplitude oscillations, the long time average agrees well with the RMT prediction in Eq.\ \eqref{eq:IPRCUE}. It is not required that we average over any particular basis in order to agree with the prediction in Eq.\ \eqref{eq:IPRCUE}, only that the basis does not introduce any special symmetries to $F_n(t)$. As a demonstration, panel (b) shows the survival probability averaged over a randomly generated complex orthonormal basis (via a QR decomposition of a randomly populated complex matrix), which we denote $\overline{P}_{\mathrm{R}}(t)$. The long-time average in the random basis is $\overline{P}_{\mathrm{R}} \approx 1.969$, and agrees extremely well (within $<0.1\%$) with $\mathrm{IPR}_{\mathrm{CUE}}|_{N=60}$. Finally, panel (c) of Fig.\ \ref{fig:PxPz} shows $P_n(t)$ averaged over the $\hat{S}_z$ basis. The long-time average is $\overline{P}_z\approx 3.229$, again shown as a red-dashed line, and has an approximately $11\%$ deviation from $\mathrm{IPR}_{\mathrm{COE}}|_{N=60}\approx  2.905$. In each image, the TTC order is $n = 40$ and the system size is $N = 60$. 

Taken together, the plots shown in Fig.\ \ref{fig:PxPz} (see especially the log-log plots in the insets) indicate that the effective dynamics of the $n$-fold TTC are regular at short times, then undergo a transition period until approximately $t_{\mathrm{Th}}$, shown in all panels as a vertical dashed line according to Eqs.\ \eqref{eq:ThouCUE}-\eqref{eq:IPRCOE}, after which the survival probability settles down, or at least oscillates about a universal value given by its RMT prediction in the respective ensemble. In general, we expect the agreement between the numerical result and the theoretical predictions given in Eqns.\ \eqref{eq:IPRCUE} and \eqref{eq:IPRCOE}  to improve for higher TTC order $n$ and correspondingly longer time averages.



\section{Summary and Conclusion}
\label{Sec:Conclusion}

In this paper, we have given a proof-of-concept that conventional signals of chaotic behaviour, including level statistics, classical Lyapunov exponents, and RMT predictions can arise from generalized $n$-fold TTCs even when starting from a non-chaotic Hamiltonian.  This is significant because TTCs, and more specifically OTOCs (the $n=2$ case), are commonly used as diagnostic tools for what is often called quantum chaos.  Higher-order TTCs have been the focus of some previous studies \cite{bhattacharyya19,haehl19,halpern18,shenker14,roberts17}, including as an improvement of the standard four-point OTOC as a diagnostic of chaos, and so the emergence of signatures of chaos as a result of the form of the TTC is therefore of interest. 

In our case we chose a rather simple system consisting of a bosonic Josephson junction coupled dispersively to an AQD or impurity spin. This has the benefit of being integrable and thus all the signals of chaos we find are genuinely in the TTC not the original dynamics. Importantly, both the parameters $\alpha_x$ and $\alpha_z$ must be nonzero in the system we are considering so that there are noncommuting pieces in the Hamiltonian.  In particular, $\alpha_x=0$ would allow the vanishing of the commutator $[ \hat{H}_1, \hat{H}_2  ] = 0$, resulting in trivial dynamics, while $\alpha_z$ is necessary to destroy the parity symmetry and thereby allow the eigenstates of $\hat{\mathcal{F}}$ and $\hat{U}_t$ to be modelled by the appropriate ensembles in RMT.  However, the simplicity of our system also means that even the effectively shaken $n$-fold TTC is only weakly chaotic and thus our survival probabilities do not rapidly converge to the RMT values (for our parameters the deviation can be as large as $11\%$). However, we also saw evidence that the bases of $\hat{S}_{x}$ and $\hat{S}_{z}$ remain in some sense special (being the eigenstates of the two pieces of the Hamiltonian) whereas when a truly random basis was chosen we obtained excellent agreement with deviation of less than $0.1\%$.

There remain signatures of chaos which we have not addressed here. For example, we have not attempted to identify a \textit{quantum} Lyapunov exponent  (one typically uses $F_2(t)$ and finds that $F_2(t) \approx 1 -c e^{\lambda t}$ \cite{shen17}, where $c$ is some constant). Rather, in Sec.\ \ref{Sec:classical} we have merely looked directly at the classical Lyapunov exponent obtained from the mean-field equations of motion for the $n$-fold TTC. Furthermore, for a chaotic system obeying the RMT predictions outlined in this paper, it is expected that there exists a `correlation hole' (a signature of correlations in level statistics) in the survival probability with a minimum at $t_{\mathrm{Th}}$ \cite{pechukas84,leviandier86,alhassid92,schiulaz19,hererra19}, which proceeds to ramp towards the saturation value. We have not identified a correlation hole in the TTCs studied here, likely because the transition from a regular to chaotic system occurs explicitly in time, so level correlations will not be detectable prior to $t_{\mathrm{Th}}$.

\acknowledgments  WK is grateful to Greg Kaplanek for useful discussions. We acknowledge the Natural Sciences and Engineering Research Council of Canada (NSERC) for funding.

\appendix

\section{TTC Calculation}
\label{app:TTCCalc}

We consider the two-time correlator,
\begin{equation}
	F_n(t)=\Bigl\langle \left[\hat{A}(t)\hat{B}(0) \right]^{n} \Bigr\rangle
\end{equation}
where $\hat{A}(t)=\mathrm{e}^{\mathrm{i}\hat{H}t}\hat{A}\mathrm{e}^{-\mathrm{i}\hat{H}t}$ and assume that $\hat{A}$ and $\hat{B}$ initially commute. Selecting $\hat{A}=\hat{B}=\hat{\sigma}_x$, that is, the hopping operator on the Hilbert space of the quantum dot,
\begin{widetext}
	\begin{align}
		F_n(t)=&\;\biggl\langle\biggl\{\exp\left[\mathrm{i}\left(\frac{k_z}{N+1}\hat{S}_z^2-\alpha_x\hat{S}_x+\alpha_z\hat{S}_z-\frac{\Delta}{2}(\mathds{1}+\hat{\sigma}_z)+\beta\hat{S}_x(\mathds{1}+\hat{\sigma}_z)\right)t\right]\\&\;\times \hat{\sigma}_x\exp\left[-\mathrm{i}\left(\frac{k_z}{N+1}\hat{S}_z^2-\alpha_x\hat{S}_x+\alpha_z\hat{S}_z-\frac{\Delta}{2}(\mathds{1}+\hat{\sigma}_z)+\beta\hat{S}_x(\mathds{1}+\hat{\sigma}_z)\right)t\right]\hat{\sigma}_x\biggr\}^{n}\biggr\rangle\nonumber\\
		=&\;\biggl\langle\biggl\{\exp\left[\mathrm{i}\left(\frac{k_z}{N+1}\hat{S}_z^2-\alpha_x\hat{S}_x+\alpha_z\hat{S}_z-\frac{\Delta}{2}(\mathds{1}+\hat{\sigma}_z)+\beta\hat{S}_x(\mathds{1}+\hat{\sigma}_z)\right)t\right] \label{eq:appendixFn} \\&\;\times \exp\left[-\mathrm{i}\left(\frac{k_z}{N+1}\hat{S}_z^2-\alpha_x\hat{S}_x+\alpha_z\hat{S}_z-\frac{\Delta}{2}(\mathds{1}-\hat{\sigma}_z)+\beta\hat{S}_x(\mathds{1}-\hat{\sigma}_z)\right)t\right]\biggr\}^{n}\biggr\rangle\nonumber\;,
	\end{align}
\end{widetext}
where we have made use of the fact that for some function of the Pauli spin matrices $f(\hat{\sigma}_x,\hat{\sigma}_y,\hat{\sigma}_z)$, $\hat{\sigma}_xf(\hat{\sigma}_x,\hat{\sigma}_y,\hat{\sigma}_z)\hat{\sigma}_x=f(\hat{\sigma}_x,-\hat{\sigma}_y,-\hat{\sigma}_z)$.

For the expectation value we assume a product state $\ket{\psi}_B\otimes\ket{+}_d$, where $\ket{\psi}_B$ is a general state of the BEC and $\ket{+}_d$ is the excited state of the AQD. 
The $(\mathds{1}-\hat{\sigma}_z)$ and $(\mathds{1}+\hat{\sigma}_z)$ factors in Eq.\ (\ref{eq:appendixFn}) act as projectors onto the excited and ground AQD states, respectively. In particular, the operator $(\mathds{1}-\hat{\sigma}_z)$ appearing in the right hand exponential acts as twice the identity operator on the ket $\ket{+}_d$ and thus $\ket{+}_d$ passes through this exponential replacing all the $(\mathds{1}-\hat{\sigma}_z)$ factors by the number 2.  Meanwhile, the other exponential contains $(\mathds{1}+\hat{\sigma}_z)$ which annihilates $\ket{+}_d$ and so expanding out the exponential in a Taylor series we find all the AQD operators vanish and we can re-sum the exponential with only the boson operators. As a result we find the surprising result that the correlation function becomes completely independent of the AQD and we are left with
\begin{align}
	F_n(t)=&\;\biggl\langle\biggl\{e^{\mathrm{i}\left(\frac{k_z}{N+1}\hat{S}_z^2-\alpha_x\hat{S}_x+\alpha_z\hat{S}_z\right)t } \\ & \times e^{-\mathrm{i}\left(\frac{k_z}{N+1}\hat{S}_z^2-\alpha_x\hat{S}_x+\alpha_z\hat{S}_z-\Delta+2\beta\hat{S}_x\right)t } \biggr\}^{n}\biggr\rangle_B \nonumber
\end{align}
The $\Delta$ term results in a global phase which we can choose to set to zero, and we are free to select $\beta$ as we wish; we choose $\beta=\alpha_x/2$ to remove all $\hat{S}_{x}$ terms in $\hat{H}_{2}$ as described in the main text, so that we finally achieve,
\begin{align}
	F_n(t)=&\;\biggl\langle\biggl\{\exp\left[\mathrm{i}\left(\frac{k_z}{N+1}\hat{S}_z^2-\alpha_x\hat{S}_x+\alpha_z\hat{S}_z\right)t\right]\\&\times\exp\left[-\mathrm{i}\left(\frac{k_z}{N+1}\hat{S}_z^2+\alpha_z\hat{S}_z\right)t\right]\biggr\}^{n}\biggr\rangle_B\nonumber\\
	\equiv&\;\bigl\langle\hat{\mathcal{F}}^{n}\bigr\rangle_B=\biggl\langle\biggl[\mathrm{e}^{-\mathrm{i}\hat{H}_1t}\mathrm{e}^{-\mathrm{i}\hat{H}_2t}\biggr]^{n}\biggr\rangle_B\;.
\end{align}

\section{\label{app:MeanField}Mean-Field Hamiltonian and Equations of Motion}

In order to write down the mean-field approximation to the quantum Hamiltonians given in Eqns.\ \eqref{eq:ham}, \eqref{eq:ham1}, and \eqref{eq:ham2}, we first write the spin operators in terms of their corresponding Schwinger representations,
\begin{align}
	\hat{S}_{z} = \;& (\hat{b}_{L}^{\dag}\hat{b}_{L}-\hat{b}_{R}^{\dag}\hat{b}_{R})/2 \\
	\hat{S}_{x} = \;& (\hat{b}_{R}^{\dag}\hat{b}_{L}+\hat{b}_{R}^{\dag}\hat{b}_{R})/2 \; .
\end{align}
where $\hat{b}_{L/R}^{(\dag)}$ annihilates (creates) a boson in the left/right well. The quantum dot operators can be similarly written, using $\hat{d}_{u/d}^{(\dag)}$, 
\begin{equation}
	\hat{\sigma}_z=\left(\hat{d}_u^\dagger\hat{d}_u-\hat{d}_d^\dagger\hat{d}_d\right)/2\;.
\end{equation}
Next, we assume that in the classical limit $N \rightarrow \infty$, we can replace the boson operators by complex numbers $\hat{b}_i\to \sqrt{n_i}\mathrm{e}^{\mathrm{i}\phi_i}$. We can also make a similar replacement for the dot $\hat{d}_i\to \sqrt{n_i}\mathrm{e}^{\mathrm{i}\varphi_i}$ (the ``mean-field'' theory is in fact exact for the dot), and then defining $n_L=N-n_R$, $\phi=\phi_L-\phi_R$, $z=(n_L-n_R)/N$, and $y=n_u-n_d$ we have the substitution rules,
\begin{align}
	\hat{S}_x\to&\;\sqrt{n_Ln_R} \cos \left(\phi _L-\phi _R\right)=\frac{N}{2}\sqrt{1-z^2}\cos\phi\\
	\hat{S}_z\to&\; \frac{1}{2}(n_L-n_R)=\frac{Nz}{2}\\
	\hat{\sigma}_z\to&\;\frac{1}{2}(n_u-n_d)=\frac{y}{2}\;.
\end{align}
Hence, the mean-field Hamiltonian ($\mathcal{H}=\lim_{N\to\infty}\hat{H}/N$) is,
\begin{align}
	\mathcal{H}=&\;\frac{k_z}{4}z^2-\frac{\alpha_x}{2}\sqrt{1-z^2} \cos\phi+\frac{1}{2}\alpha_zz-\frac{\Delta}{2N}\left(1+\frac{y}{2}\right)\nonumber\\&+\frac{\beta}{2}\sqrt{1-z^2} \cos\phi\left(1+\frac{y}{2}\right) \label{eq:MFHamil}
\end{align}
where all energies on the RHS are measured in terms of $\alpha_x$. Hamilton's equations give,
\begin{align}
	\dot{z}=&\;-\frac{\partial \mathcal{H}}{\partial \phi}=-\frac{\alpha_x}{2}\sqrt{1-z^2}\sin\phi \nonumber \\ & \quad +\frac{\beta}{2}\sqrt{1-z^2}\sin\phi\left(1+\frac{y}{2}\right)\\
	\dot{\phi}=&\;\frac{\partial \mathcal{H}}{\partial z}=\frac{k_z}{2}z+\frac{\alpha_z}{2}+\frac{\alpha_xz\cos\phi}{2\sqrt{1-z^2}} \nonumber \\
	& \quad -\frac{\beta z\cos\phi}{2\sqrt{1-z^2}}\left(1+\frac{y}{2}\right)\\
	\dot{y}=&\;-\frac{\partial \mathcal{H}}{\partial \varphi}=0\\
	\dot{\varphi}=&\;\frac{\partial \mathcal{H}}{\partial y}=\frac{\beta}{4}\sqrt{1-z^2}\cos\phi-\frac{\Delta}{4N}
\end{align}
Likewise, the mean-field approximations for $\hat{H}_1$ and $\hat{H}_2$ are,
\begin{align}
	\mathcal{H}_1=&\;-\frac{k_z}{4}z^2+\frac{\alpha_x}{2}\sqrt{1-z^2} \cos\phi-\frac{\alpha_z}{2}z\label{eq:HMF1}\\
	\mathcal{H}_2=&\;\frac{k_z}{4}z^2+\frac{\alpha_z}{2}z\label{eq:HMF2}\;.
\end{align}
The classical trajectories for Eqs.\ \eqref{eq:HMF1}-\eqref{eq:HMF2} are similarly calculated using Hamilton's equations, however the set of conjugate variables $\{y,\varphi\}$ is no longer present, 
\begin{align}
	\dot{z}_1=&\;-\frac{\alpha_x}{2}\sqrt{1-z^2}\sin\phi\\
	\dot{\phi}_1=&\;-\frac{k_zz}{2}-\frac{\alpha_z}{2}-\frac{\alpha_xz\cos\phi}{2\sqrt{1-z^2}}\\
	\dot{z}_2=&\;0\\
	\dot{\phi}_2=&\;\frac{k_zz}{2}+\frac{\alpha_z}{2}
\end{align}
The dynamics are governed by repeatedly alternating between $\mathcal{H}_1$ and $\mathcal{H}_2$ for a time $\alpha_x T$.
\begin{figure}[t!]
	\includegraphics[width=\columnwidth]{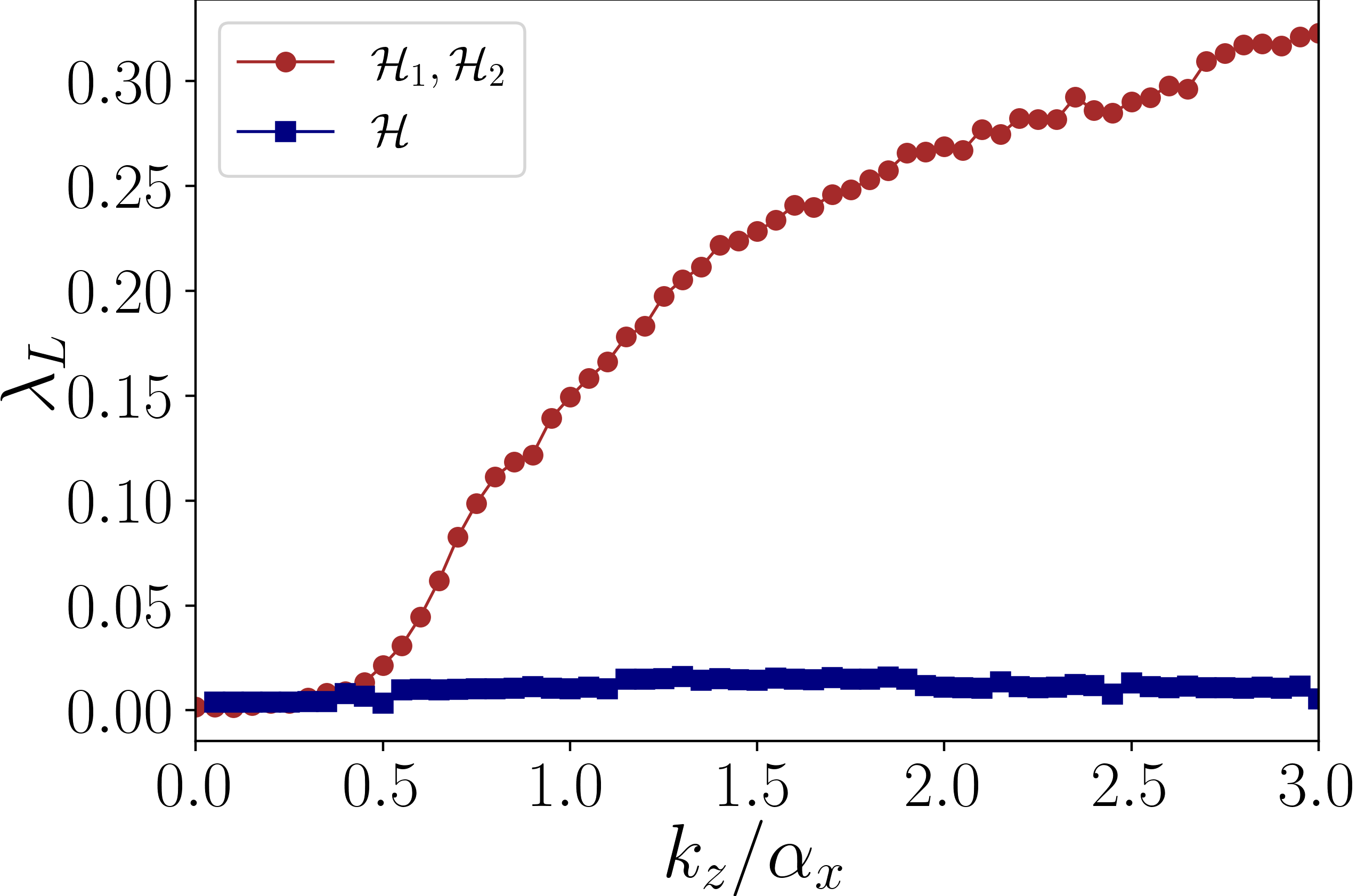}
	\caption{The Lyapunov exponent, $\lambda_L$, as a function of $k_z$ for dynamics produced by the mean-field version of Eq.\ \eqref{eq:ham} (blue squares) and the mean-field versions of Eqns.\ \eqref{eq:ham1} and \eqref{eq:ham2} (red dots).  We can see for $\mathcal{H}$, $\lambda_L \approx 0$ suggesting the system is regular while for the combination of $\mathcal{H}_1$ and $\mathcal{H}_2$, $\lambda_L > 0$ suggesting chaotic dynamics.  Each data point is the maximum $\lambda_L$ averaged over 1500 random initial states in phase space.  For both sets of data $\alpha_z = 0.01 \alpha_x$ and for the red data the dynamics is cycled through $\mathcal{H}_1$ and $\mathcal{H}_2$ $n = 20$ times.}
	\label{fig:LEvsk}
\end{figure}

In Fig.\ \ref{fig:LEPic} of the main text, we show the effects of tuning the time at which the TTC is measured on the Lyapunov exponent for the classical Hamiltonians \eqref{eq:HMF1}-\eqref{eq:HMF2}. Alternatively, in Fig.\ \ref{fig:LEvsk}, we demonstrate the presence of a positive classical Lyapunov exponent for the system alternating between $\mathcal{H}_1$ and $\mathcal{H}_2$ as a function of the bosonic interaction energy. The original Hamiltonian has no appreciable Lyapunov exponent while the shaken system rapidly develops exponential separation of trajectories after $k_z/\alpha_x>0.5$. The distance $d$ between trajectories $\{z_n,\phi_n\}$ and $\{z_m,\phi_m\}$ in the BEC coordinates corresponds to a great-circle distance on the Bloch sphere,
\begin{equation}
	d=\cos^{-1}\left[z_nz_m+\sqrt{(1-z_n^2)(1-z_m^2)}\cos(\phi_n-\phi_m)\right]\;,
\end{equation}
which can be used to calculate $\lambda_L$.

\section{Individual Survival Probability}
\label{app:SurvProb}

In Sec.\ \ref{SubSec:Survival}  of the main text we concerned ourselves mainly with the general features of the survival probability without specifically selecting states. The qualitative nature of $P_n(t)$ broadly follows $\overline{P}_n(t)$ (the average), although it is more sensitive due to the lack of averaging over initial states. The saturation values when not averaged over the bases are then,
\begin{align}
	P^{\mathrm{CUE}}=&\;\frac{2}{\mathcal{D}+1}\label{eq:PCUE}\\
	P^{\mathrm{COE}}=&\;\frac{3}{\mathcal{D}+2}\label{eq:PCOE}\;.
\end{align}

\begin{figure}[t!]
	\includegraphics[width=\columnwidth]{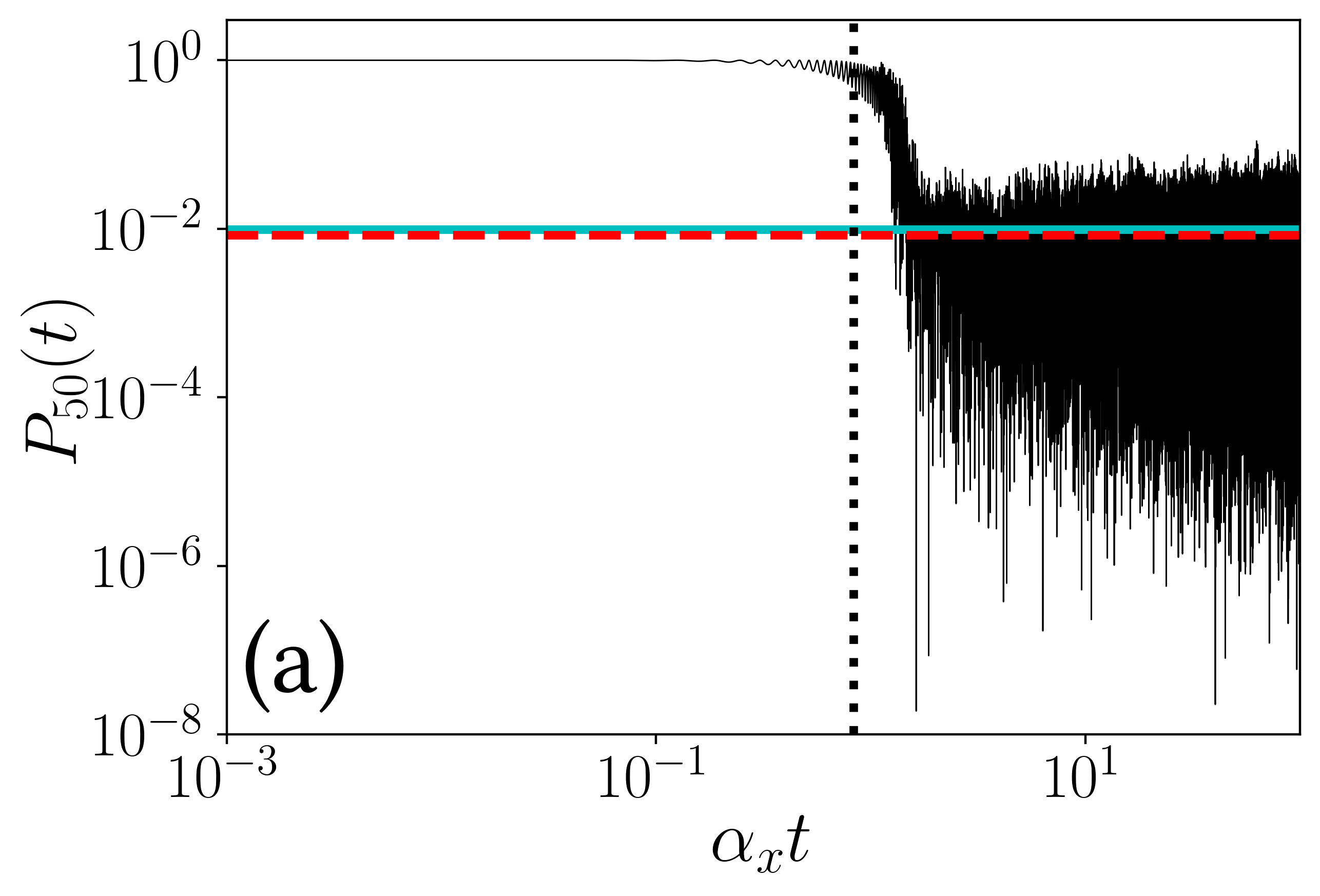}
	\includegraphics[width=\columnwidth]{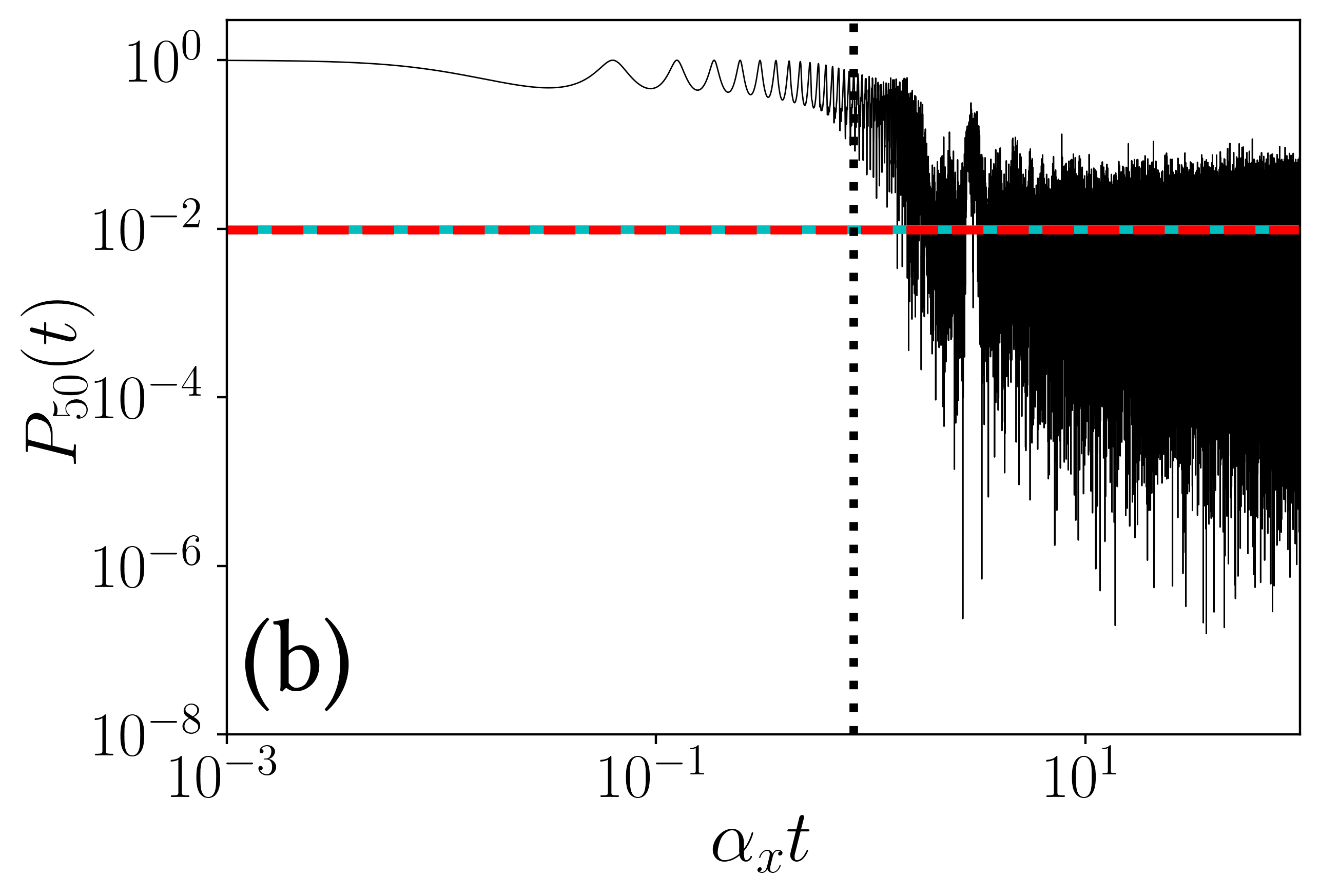}
	\includegraphics[width=\columnwidth]{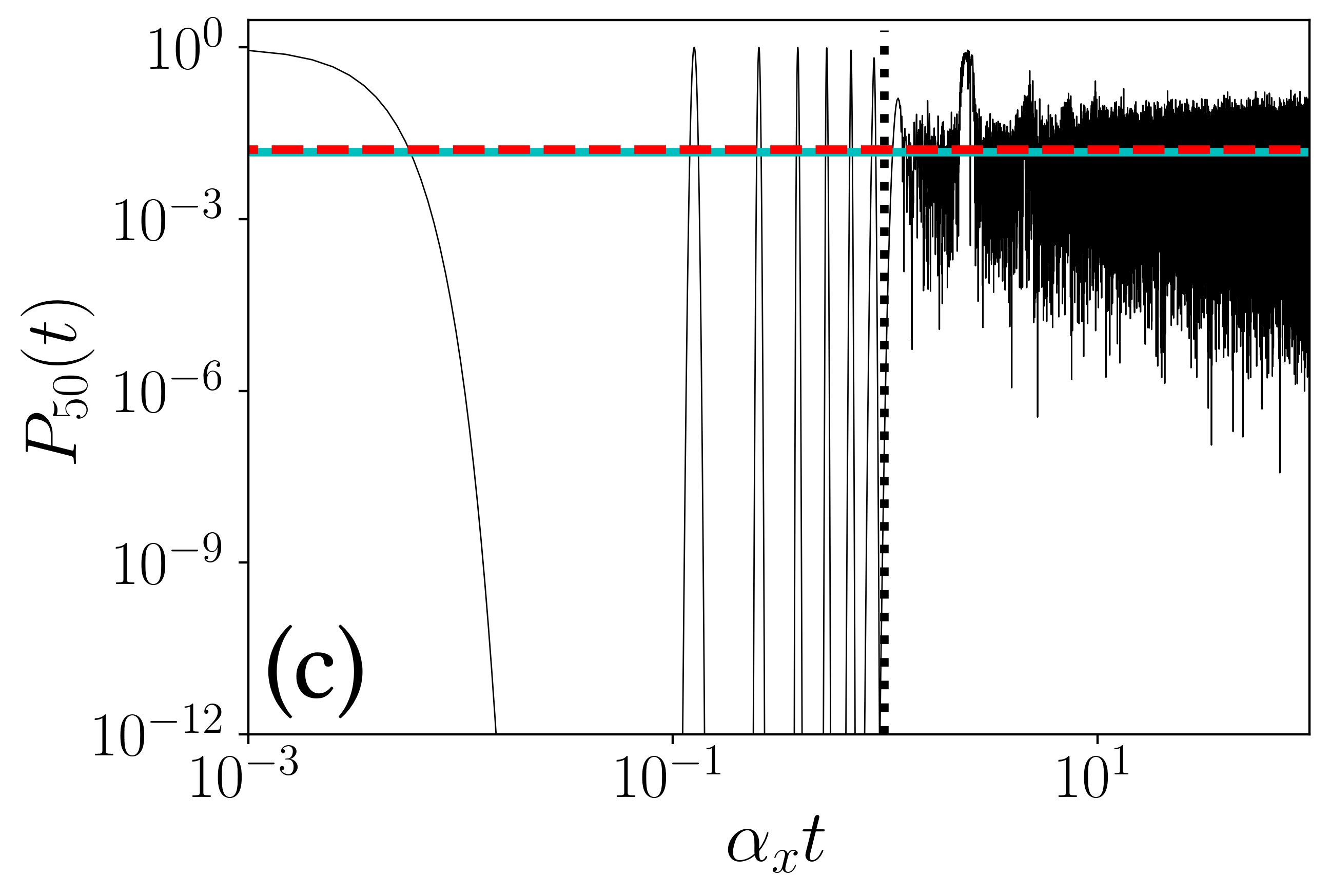}
	\caption{Survival probability $P_n(t)$ for two different choices of $\ket{\psi}$. \textbf{Panel (a):} Using $\ket{\psi}=\ket{N/2}_x=\mathrm{e}^{-\mathrm{i}\hat{S}_y\pi/2}\ket{N/2}$ \textbf{Panel (b):} Using the Gaussian state given in Eq. \eqref{eq:GaussianState} \textbf{Panel (c):} Using $\ket{\psi}=\ket{N/2}$. $t_{\mathrm{Th}}$ is shown as a vertical dotted line, the long-time average of the data is given as a horizontal solid cyan line and the corresponding RMT prediction is given by a dashed red line. The parameter values in each image is $k_z=3\alpha_x$, $\alpha_z=0.01\alpha_x$, $N=200$ and $n=50$.}
	\label{fig:Survival}
\end{figure}

In Fig. \ref{fig:Survival}, we show the survival probability $P_n(t)$ for three different states using $n=50$ and $N=200$ at $k_z=3\alpha_x$ and $\alpha_z=0.01\alpha_x$. Panel (a) shows the survival probability for $\ket{\psi}=\ket{N/2}_x=\mathrm{e}^{-\mathrm{i}\hat{S}_y\pi/2}\ket{N/2}$, which is the ground state of the $\hat{S}_x$ operator and also a coherent state. Much like the average, the survival probability is roughly constant for short times, then drops off, saturating at approximately $8.521\times 10^{-3}$, which is within $14\%$ of the CUE value of $9.901\times 10^{-3}$. For panel (b), we chose an initial state which is Gaussian (but not a coherent state) in the $\hat{S}_z$ basis,
\begin{equation}\label{eq:GaussianState}
	\ket{\psi}=\frac{1}{(2\pi N)^{1/4}}\sum_{m=-N/2}^{N/2}\mathrm{e}^{-\frac{m^2}{4N}}\ket{m}
\end{equation}
This broader state has excellent agreement with the CUE value at approximately $9.757\times 10^{-3}$, within $1.5\%$. Finally, in panel (c), we instead choose a member of the $\hat{S}_z$ basis, $\ket{\psi}=\ket{N/2}$. The return probability of this state demonstrates sharp peaks reminiscent of dynamical phase transitions (DPT) and subsequently saturates at $16.605\times 10^{-3}$, approximately $12\%$ higher than $P^{\mathrm{COE}}$.

Comparing the actual saturation values with those predicted in Eqs. \eqref{eq:PCUE}-\eqref{eq:PCOE}, we find that the relative errors can be fairly large, on the order of 10-15$\%$. However, the errors on individual states can be extremely sensitive to parameter shifts. For example, a change to $k_z=8\alpha_x$ (deeper into the `chaotic' region, extrapolated from Fig.\ \ref{fig:LEvsk}) can reduce the error from the coherent state ($\ket{\psi}=\ket{N/2}_x$) to approximately $9.393\times 10^{-3}$, that is $5.1\%$ error from the $P^{\mathrm{COE}}$ value. 


The return probability can occasionally reach extremely small orders of magnitude, especially for larger $N$, at which machine-precision exact diagonalization becomes insufficient to properly resolve $P_n(t)$. This effect occurs in the regular regions prior to $t_{\mathrm{Th}}$ in panel (c) of Fig.\  \ref{fig:Survival}, where DPT-like sharp valleys can only be resolved with precision on the order of $80$ decimal places (increasing with $N$). The extreme sensitivity of numerical noise in the Loschmidt echo in similar systems has been identified in Ref. \cite{homrighausen17}. The region which requires high sensitivity to properly resolve the dynamics is, however, not our primary concern since it is not the chaotic region.

\end{document}